\documentclass[aps,nofootinbib,prx,twocolumn,notitlepage,floatfix,superscriptaddress,10pt,tightenlines]{revtex4-2}

\usepackage{amsmath,amssymb,mathtools}
\usepackage{graphicx}
\usepackage{physics}        
\usepackage{revsymb4-2}     
\usepackage{xcolor}
\usepackage{comment}
\usepackage{hyperref}
\usepackage{soul}
\hypersetup{colorlinks=true, linkcolor=blue, citecolor=blue, urlcolor=blue}

\begin{document}

\title{Learning parameter curves in feedback-based quantum algorithms}

\author{Vicente Pe\~na P\'erez}
\email{vpenaper@asu.edu}
\affiliation{School of Electrical, Computer, and Energy Engineering, Arizona State University, Tempe, AZ 85287, USA}
\affiliation{Quantum Algorithms and Applications Collaboratory, Sandia National Laboratories, Albuquerque, NM 87185, USA}

\author{Matthew D. Grace}
\affiliation{Quantum Algorithms and Applications Collaboratory, Sandia National Laboratories, Albuquerque, NM 87185, USA}

\author{Christian Arenz}
\email{carenz1@asu.edu}
\affiliation{School of Electrical, Computer, and Energy Engineering, Arizona State University, Tempe, AZ 85287, USA}

\author{Alicia B. Magann}
\email{abmagan@sandia.gov}
\affiliation{Quantum Algorithms and Applications Collaboratory, Sandia National Laboratories, Albuquerque, NM 87185, USA}

\date{\today}

\begin{abstract}
Feedback-based quantum algorithms (FQAs) operate by iteratively growing a quantum circuit to optimize a given task. At each step, feedback from qubit measurements is used to inform the next quantum circuit update.     
In practice, the sampling cost associated with these measurements can be significant. Here, we ask whether FQA parameter sequences can be predicted using classical machine learning, obviating the need for qubit measurements altogether. To this end, we train a teacher–student model to map problem instances to associated FQA parameter curves in a single classical inference step. We first study weighted MaxCut problem instances, where numerical experiments show that this machine learning model can accurately predict FQA parameter curves across a range of problem sizes, including sizes not seen during model training. We then extend the analysis to transverse-field Ising models on graphs, and analyze how this learning strategy performs and generalizes beyond the original MaxCut setting. To evaluate performance, we compare the predicted parameter curves in simulation against FQA reference curves and linear quantum annealing schedules. We observe similar results to the former and performance improvements over the latter. These results suggest that machine learning can offer a heuristic, practical path to reducing sampling costs and resource overheads in quantum algorithms.
\end{abstract}

\maketitle

\section{Introduction}

There is substantial interest in using quantum computers to approximate ground states of Hamiltonians. This is a challenging computational task \cite{doi:10.1137/S0097539704445226,PhysRevLett.102.130503} with diverse applications, including in combinatorial optimization \cite{lucas14}. Early work in the latter domain centered on quantum annealing (QA) \cite{kadowaki98, farhi00, das08, adiabatic3}, a conventionally continuous-time approach that aims to prepare the ground state of an Ising Hamiltonian using an adiabatic evolution, parameterized according to an annealing schedule. Later, variational quantum algorithms (VQAs) emerged as another approach for ground state preparation geared towards implementations on digital, gate-model quantum computers \cite{cerezo21,RevModPhys.94.015004}. These algorithms operate by classically optimizing over a set of free parameters in a quantum circuit, whose purpose is to prepare an approximation of the target ground state. One notable example of a VQA is the quantum approximate optimization algorithm (QAOA) \cite{farhi14}, developed in analogy to QA for finding approximate solutions to combinatorial optimization problems. 

Feedback-based quantum algorithms, such as the feedback-based algorithm for quantum optimization (FALQON) \cite{magann22, magann22-2}, were subsequently developed in analogy to QAOA.  These algorithms operate by assigning values to free parameters in quantum circuits in a sequential manner. At each step in the sequence, a feedback law is utilized to set the value of a given parameter according to the outcome of qubit measurements performed at the previous step. Then, additional qubit measurements are performed to inform the next step. The feedback law is derived from quantum Lyapunov control principles \cite{Grivopoulos9,Cong1} to ensure that a cost function decreases monotonically as a function of step. In practice, however, the sampling cost associated with evaluating the feedback law can be significant, as each step requires many circuit repetitions to perform the needed qubit measurements. This sampling overhead can become a significant bottleneck towards quantum device implementations at scale. Here, we address this challenge by leveraging classical machine learning (ML) to {predict} FALQON parameters without the need for measurement data.

\begin{figure}[t]
    \centering
    \includegraphics[width=0.9\columnwidth]{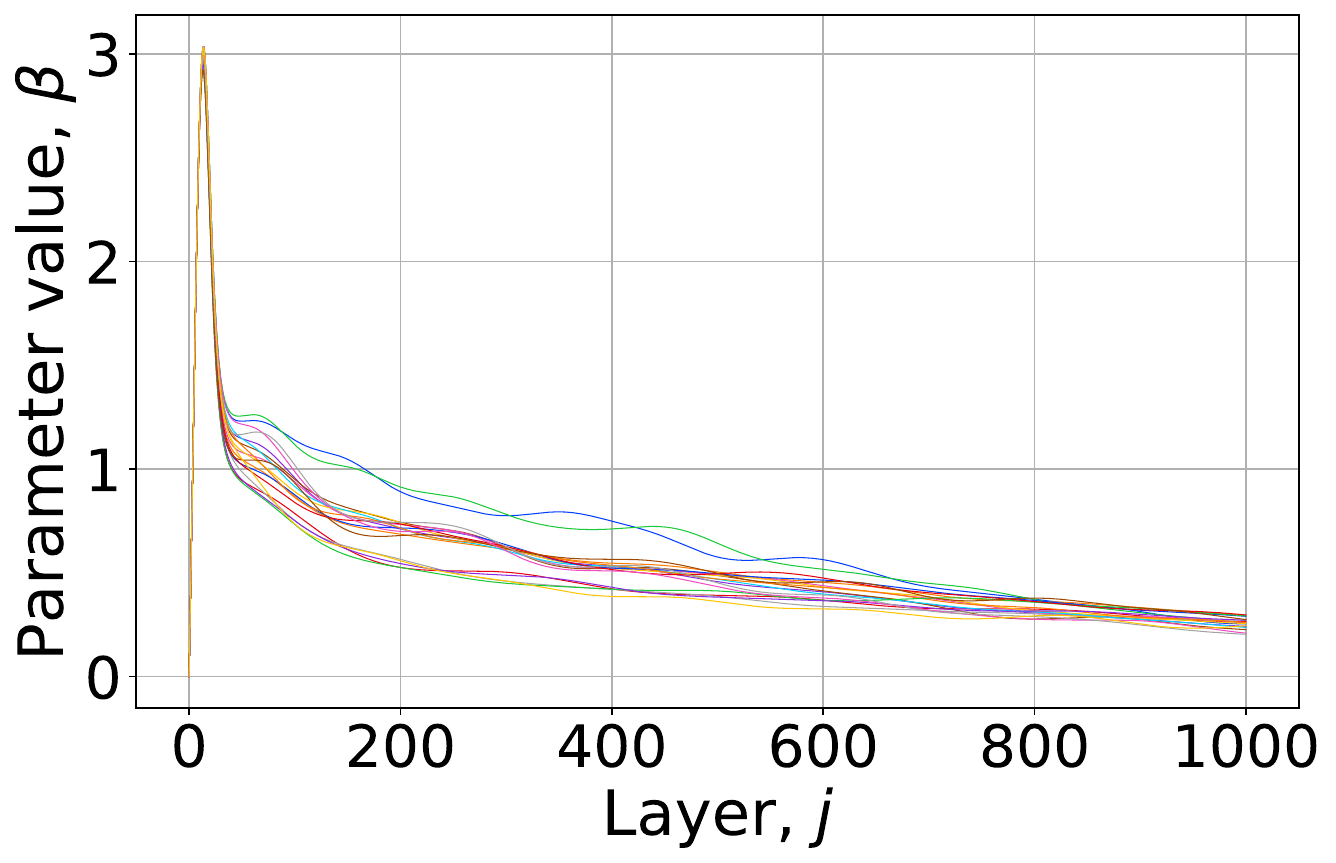}
    \caption{FALQON parameter curves found for solving MaxCut on 19 non-isomorphic, unweighted 3-regular graphs with ten nodes, where each curve corresponds to the FALQON solution for one of the 19 graphs. The consistent clustering of these curves around a common profile motivates our work exploring if machine learning could be employed to predict them.}
    \label{fig:parameter-curve}
\end{figure}

The sequence of FALQON parameters can be collected into a \emph{parameter curve}, which is analogous to parameter sequences in QAOA and continuous annealing schedules in QA. FALQON parameter curves tend to display a typical shape, characterized by an initial {peak region} defined by large, rapid variations in the FQA parameters, followed by a subsequent {tail region}. Figure~\ref{fig:parameter-curve} illustrates this typical shape for FALQON parameter curves associated with the MaxCut problem. Across different MaxCut problem instances, the curves cluster tightly around the common profile. 
This phenomenon was observed but not analyzed in Ref. \cite{magann22}, and it suggests that FALQON parameter curves may be predictable from relatively simple problem features, without having to explicitly run the full feedback procedure for each instance.

Motivated by this observation, in this work we develop and test an ML model that replaces the per-layer qubit measurements with a one-step classical prediction of the full parameter curve. We develop this ML model in the context of predicting FQA parameter curves associated with solving an Ising model formulation of the weighted MaxCut problem, and analyze its performance. We then explore applications to quantum models, here considering transverse-field Ising models (TFIM) defined on the same class of graphs considered for MaxCut.
While our results appear encouraging, we emphasize that our ML approach is heuristic: what problem classes admit FQA parameter curves compatible with prediction using classical ML, and when ML models trained on one family of Hamiltonians can be productively transferred to others, remain open questions.

This direction is aligned with prior efforts that use ML to predict QAOA parameters, primarily to reduce the burden of classical optimization \cite{alam20,jain22,xie23,paolo24,tyagin2025qaoagptefficientgenerationadaptive}. Ultimately, the ML predicted curves we obtain could be used as {standalone} solutions that directly replace FALQON, QA, or QAOA solutions, or as {warm starts} for subsequent refinement through classical optimization or hybrid FALQON or QAOA strategies \cite{magann22-2}.

The remainder of this article is organized as follows. Section~\ref{sec:background} reviews QA, QAOA, and FALQON, contrasting feedback-generated FALQON parameter curves with those of QAOA and QA. Section~\ref{sec:model} introduces our teacher-student approach for predicting the full FALQON parameter curves, detailing the high-level architecture and training procedure. Section~\ref{sec:results} reports a variety of numerical results that explore the performance of the ML predictions. Section~\ref{sec:outlook} summarizes and outlines future directions. 

\section{Background}
\label{sec:background}

In this section, we provide relevant technical background on QA and QAOA, followed by an overview of FALQON. 

\subsection{Quantum annealing}\label{sec:qa}

Quantum annealing (QA) is a strategy for preparing or approximating the ground state of a target problem Hamiltonian, $H_{\text{p}}$ \cite{kadowaki98, farhi00, das08, adiabatic3}. It operates by first initializing a set of qubits in the ground state of another, simpler Hamiltonian, denoted here by $H_{\text{d}}$. Then, the qubits are evolved over time under a Hamiltonian,
\begin{equation}
    H(t) = A(t)\,H_{\text{d}} + B(t)\,H_{\text{p}},\,t\in[0,T]
    \label{eq:qa_hamiltonian}
\end{equation}
that smoothly interpolates between the simpler Hamiltonian and the problem Hamiltonian of interest, according to a predefined annealing {schedule} captured by the time-dependent scalar functions $A(t)$ and $B(t)$, typically selected so that $A(0)=B(T)=1$ and $A(T)=B(0)=0$. The performance of QA depends strongly on the choice of QA schedule, and a commonly used heuristic choice is a linear schedule, given by
\begin{equation}
    A(t) = 1 - \frac{t}{T},
    \qquad
    B(t) = \frac{t}{T},
    \label{eq:qa_linear_schedule}
\end{equation}
which interpolates from $H_{\text{d}}$ at $t=0$ to the problem Hamiltonian $H_{\text{p}}$ at $t=T$. 
While Eq.~\eqref{eq:qa_linear_schedule} is a simple and commonly used schedule, it is not generally optimal. If some coarse-grained information about the instantaneous spectral gap is available, then the schedule can be modified to slow down near small-gap regions and speed up where the gap is larger. Such local-adiabatic strategies can improve over a purely linear interpolation, although they generally require additional spectral information about the problem Hamiltonian \cite{roland2002quantum}.
If the evolution proceeds slowly enough, the system remains close to the instantaneous ground state and transitions towards the target ground state according to the quantum adiabatic theorem \cite{born1928beweis}. Quantum annealing is conventionally an analog procedure, carried out continuously in time. In practice, the slowly varying schedules required to maintain adiabaticity can demand evolution times that are prohibitively long \cite{PhysRevLett.130.140601,garcia2025tighter}.  

It is also possible to digitize QA for implementation on gate-based quantum computers \cite{mbeng19,PhysRevLett.131.060602}. The latter has been explored in contexts such as adiabatic quantum state preparation, which has been considered as an approach for preparing initial states that are then fed into quantum algorithms such as quantum phase estimation, or algorithms for Hamiltonian simulation. Digitization can be accomplished through Hamiltonian simulation \cite{HamSim}, e.g., by discretizing the interval $[0,T]$ into $ \ell $ steps of size $\Delta t = T/\ell $ and applying a first-order Trotter formula \cite{trotter1959,suzuki1976} to yield the product 
\begin{equation}
    U_{\mathrm{QA}}(T)
    \approx
    \prod_{j=1}^{\ell}
    e^{-i \Delta t\,B(t_j) H_{\text{p}}}\,
    e^{-i \Delta t\,A(t_j) H_{\text{d}}},
    \label{eq:qa_trotter}
\end{equation}
where $t_j = j\,\Delta t$ and the error of the approximation can be controlled via the choice of $\Delta t$ according to Trotter error bounds \cite{trotter1959,suzuki1976}. Eq. (\ref{eq:qa_trotter}) describes a unitary sequence that can be used for a digitized implementation. In practice, this would be further compiled down to a gate sequence. Due to the long times that are often required for QA, such digital formulations can involve circuit depths that are substantial. 

\subsection{Quantum approximate optimization algorithm}

More recently, variational quantum algorithms (VQAs) have emerged as another paradigm for ground state preparation on digital, gate-model quantum computers \cite{cerezo21,RevModPhys.94.015004}. The idea is to execute short parameterized quantum circuits on a quantum computer, with the aim that these circuits prepare the ground state of a target Hamiltonian, or close to it. This goal is quantified using a cost function, which is typically chosen to be the target Hamiltonian expectation value, as this is minimized by the ground state. Then, a classical computer is used to iteratively search for the values of the quantum circuit parameters that minimize the cost function. Depending on the problem at hand, the resulting classical optimization landscape can be highly non-convex and can exhibit many local minima, making convergence challenging when not enough variational parameters are used \cite{wiedmann2025}. Moreover, variational training can suffer from barren plateaus, where gradients vanish exponentially with system size, rendering gradient estimation itself resource intensive \cite{McClean_Boixo_Smelyanskiy_Babbush_Neven_2018, Cerezo_2021}. 

One relevant example of a VQA is the quantum approximate optimization algorithm (QAOA) \cite{farhi14,hadfield17}, developed for finding approximate solutions to combinatorial optimization problems. The roots of QAOA can be traced back to QA. In particular, QAOA utilizes a parameterized quantum circuit {ansatz} whose structure matches that of a digitized simulation of QA in Eq. (\ref{eq:qa_trotter}). In particular, a depth-$p$ QAOA circuit prepares the quantum state
\begin{equation}
    \ket{\psi_p(\boldsymbol\gamma,\boldsymbol\beta)} \;=\;
    \Biggl(\prod_{j=1}^{p}
        e^{-i \beta_j H_{\text{d}}}\,e^{-i \gamma_j H_{\text{p}}}
    \Biggr)\ket{\psi_0},
    \label{eq:qaoa_state}
\end{equation}
where $\boldsymbol\gamma=(\gamma_1,\dots,\gamma_p)$ and
$\boldsymbol\beta=(\beta_1,\dots,\beta_p)$ are free parameters and $\ket{\psi_{0}}$ is the initial state~typically taken to be an eigenstate of $H_{\text{d}}$. That is, rather than using a fixed QA schedule, QAOA allows the digitized evolutions to be freely parameterized. The parameter values are iteratively adjusted by a classical optimizer to minimize a cost function, $\bra{\psi_p(\boldsymbol\gamma,\boldsymbol\beta)} H_{\text{p}}
    \ket{\psi_p(\boldsymbol\gamma,\boldsymbol\beta)}$, given by the expectation value of the problem Hamiltonian. This cost function is minimized by the ground state. The properties of optimal parameter curves in QAOA and QA have been analyzed previously in  \cite{PhysRevX.7.021027,brady21b,brady2021behavioranalogquantumalgorithms}.

The classical optimization loop in QAOA allows for going beyond slow, predefined QA schedules; instead, in QAOA the free parameters are iteratively optimized in an effort to achieve faster convergence than adiabatic QA or straightforward digitizations of QA, using quantum circuits with a relatively shallow depth. Nonetheless, as the size and complexity of the combinatorial problems of interest increase, the classical optimization cost in QAOA associated with searching through the associated parameter space can become a bottleneck.  A variety of proposals have aimed to address this situation, e.g., through the use of bootstrapping solutions \cite{brandao2018fixedcontrolparametersquantum, alam20, xie23}, warm starting \cite{egger2021warmstart, sack21} the classical optimization problem, or utilizing adaptively constructed circuits \cite{PhysRevResearch.4.033029, PhysRevResearch.5.033227, PhysRevA.107.062421,x8g1-7h1k,malvetti2025, mcmahon2025}.  

\subsection{Feedback-based algorithm for quantum optimization}

Feedback-based quantum algorithms constitute another approach for ground state preparation \cite{magann22, magann22-2, CLAUSEN20235171,larsen24, Rahman2024, malla2024, snht-7jsf, PhysRevResearch.7.013035, qc91-5mj2, Abdul_Rahman_2026}. Here, we consider the feedback-based algorithm for quantum optimization (FALQON), which was developed in close analogy to QAOA. In the following, we provide details on FALQON, beginning by motivating the relevant background in continuous-time quantum Lyapunov control theory \cite{Kosloff1,Sugawara2,SUGAWARA3,OHTSUKI4,Sugawara5,Mirrahimi6,Tannor7,Engel8,Grivopoulos9}. 

Consider a quantum system whose dynamics (in units of $\hbar=1$) are described by the Schrödinger equation
\begin{equation}
i\frac{d}{dt}\ket{\psi(t)}
=\bigl[\,H_{\text{p}} +\beta(t)\,H_{\text{d}}\bigr]\ket{\psi(t)},
\label{eq:continuousSch}
\end{equation}
where $|\psi(t)\rangle$ denotes the state of the system at time $t$, $H_{\text{p}}$ denotes here the portion of the Hamiltonian that is time independent, and $H_{\text{d}}$ denotes the so-called driver Hamiltonian that couples a time-dependent control function $\beta(t)$ to the system. 

Quantum Lyapunov control is a technique that can be applied to a variety of quantum control problems. Here, we consider the problem of designing a control function, $\beta(t)$, for controlling, and specifically minimizing, the expectation value of an observable, taken here to be $H_{\text{p}}$. For this application, quantum Lyapunov control operates by seeking $\beta(t)$ to satisfy the following derivative condition:
\begin{equation}
    \frac{d}{dt} \langle H_{\text{p}} \rangle_t \leq 0,
    \label{eq:derivativecondition}
\end{equation}
and in so doing, ensure that $\langle H_{\text{p}} \rangle_t$ decreases monotonically over time. Here, we have used the abbreviated notation $\langle\cdot\rangle_t \equiv \langle\psi(t)|\cdot|\psi(t)\rangle$ for the expectation value. Evaluating the left side of Eq.~\eqref{eq:derivativecondition} leads to
\begin{equation}
\frac{d}{dt}\langle H_{\text{p}}\rangle_t
=i \Bigl\langle \bigl[H_{\text{d}},H_{\text{p}}\bigr]\Bigr\rangle_t\,\beta(t)
\;=\;A(t)\,\beta(t),
\label{eq:derivativecondevaluated}
\end{equation}
with
\begin{equation}
A(t) \equiv i\langle \bigl[H_{\text{d}},H_{\text{p}}\bigr]\rangle_t.
\label{eq:continuousA}
\end{equation}
There are many choices of $\beta(t)$ that satisfy Eq.~\eqref{eq:derivativecondition} \cite{Cong1}. A common choice is to utilize a quantum Lyapunov control law of the form 
\begin{equation}
\beta(t)=- A(t)
\label{eq:feedbackLaw}
\end{equation}
to ensure that the derivative condition in Eq.~\eqref{eq:derivativecondition} is satisfied at all times $t$. This monotonic descent alone does not guarantee eventual convergence to the ground state of $H_{\text{p}}$, since the equality in Eq.~\eqref{eq:derivativecondition} is obtained whenever $A(t) = 0$ (see Ref. \cite{magann22-2}). This means that in practice, quantum Lyapunov control is often utilized as a heuristic control approach. Numerical evidence supports that it can often achieve good convergence in practice \cite{magann22, magann22-2}, with performance depending on the choice of initial state $|\psi(0)\rangle$, specification of Hamiltonians $H_{\text{p}}$ and $H_{\text{d}}$, and the functional form and parameters in the control law in Eq.~\eqref{eq:feedbackLaw}.

We now describe how digitizing this continuous-time quantum Lyapunov control approach leads to FALQON as a quantum circuit-level algorithm. This mirrors the discussion of digitizing QA in Sec. \ref{sec:qa}. First, we discretize time into steps of length $\Delta t$ and take $\beta(t)$ to be constant over each step, and second, we Trotterize the time evolution over a single time step, giving
\begin{equation}
\ket{\psi_j}
=e^{-i\,\Delta t\,H_{\text{d}}\,\beta_j}\;e^{-i\,\Delta t\,H_{\text{p}}}\;\ket{\psi_{j-1}},
\label{eq:discreteUpdate}
\end{equation}
where $\beta_j$ is the sampled control at $t_j=j\,\Delta t$,  $|\psi_j\rangle$ denotes the associated state and $j=1,2,\cdots$ labels the step or layer. The choice of $\Delta t$ has a large impact on the algorithm execution. Selecting this parameter to be too large can lead to violations of the derivative condition in Eq. (\ref{eq:derivativecondition}), while selecting it to be very small can lead to slow convergence as a function of layer. In practice, $\Delta t$ can be treated as a hyperparameter; it is desirable to tune it to be as large as possible while still satisfying Eq. (\ref{eq:derivativecondition}). 

In many cases, the digitized form in Eq.~\eqref{eq:discreteUpdate} is readily compatible with a quantum circuit-level decomposition, and it represents one step of FALQON. In other cases, e.g., where $H_{\text{p}}$ and $H_{\text{d}}$ contain multiple non-commuting terms, we can perform a subsequent Trotterization of each exponential in Eq.~\eqref{eq:discreteUpdate} to obtain a form that is then decomposable into a quantum circuit \cite{larsen24}. 

\begin{figure*}[t]
\centering
\includegraphics[width=\linewidth]{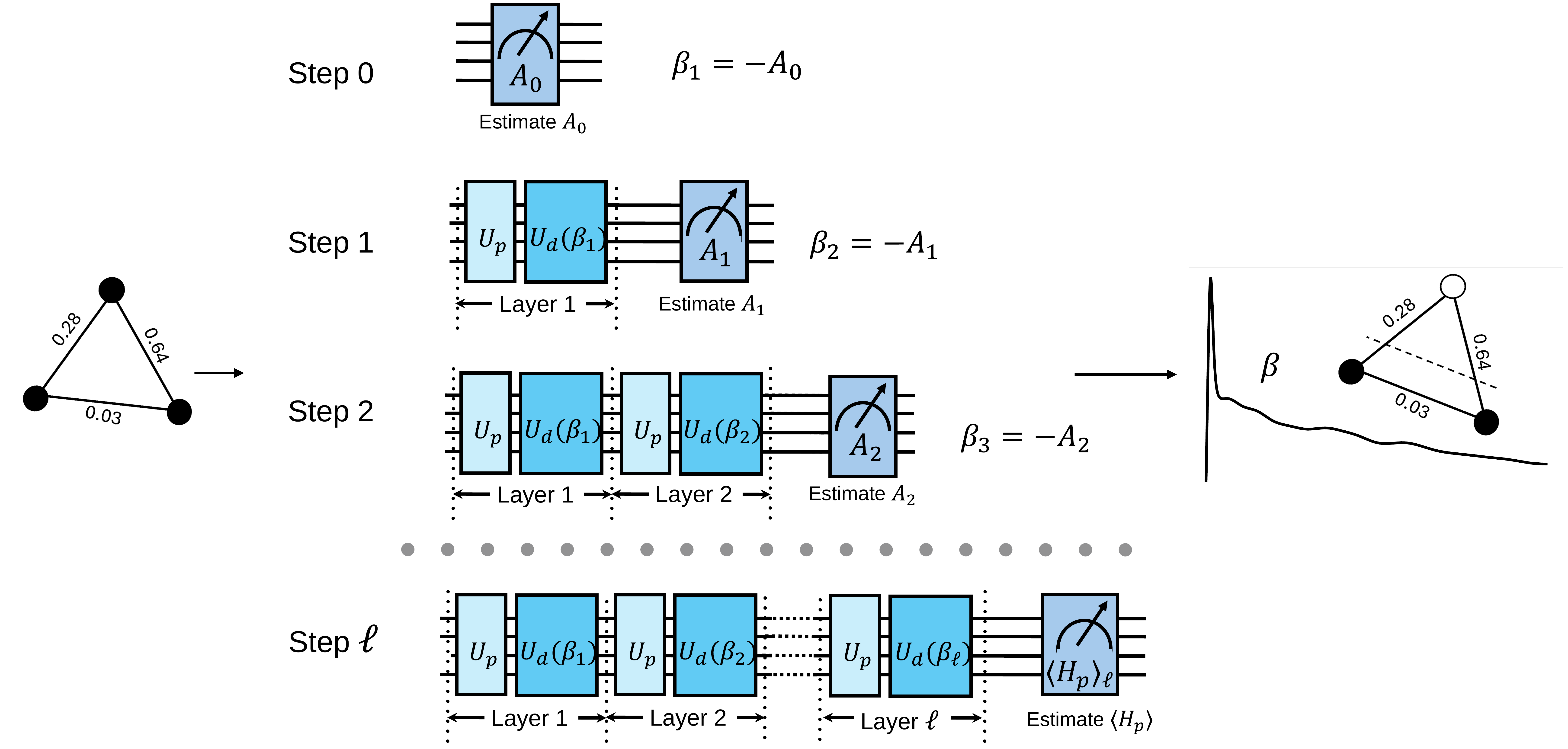}
\caption{Schematic of the steps for implementing FALQON. A weighted graph $G$ (left) defines the problem Hamiltonian $H_{\text{p}}$. Each quantum circuit layer $j$ applies $U_p=e^{-i\Delta t\,H_{\text{p}}}$ followed by $U_d(\beta_j)=e^{-i\Delta t\,\beta_j H_{\text{d}}}$. After layer $j$, the expectation value $A_j \equiv \langle i[H_{\text{d}},H_{\text{p}}]\rangle_j$ is estimated from qubit measurements and used to set the next parameter via the feedback law $\beta_{j+1}=-A_j$. Repeating for $\ell$ layers generates the parameter curve $\{\beta_j\}_{j=1}^{\ell}$, presented at right. At the end of the protocol, one may either estimate the final expectation value $\langle H_{\text{p}}\rangle_{\ell}$ or sample the terminal state to obtain a candidate MaxCut bit string solution. In the present work, we examine the prospect of replacing the steps of FALQON, and associated sampling costs, with an ML model trained to predict the full FALQON parameter curve. This figure is adapted from Ref. \cite{magann22}.}

\label{fig:FALQON protocol.}
\end{figure*}

At each step of FALQON, $\beta_j$ is determined using the quantum Lyapunov control law from Eq.~\eqref{eq:feedbackLaw}. In FALQON this control law is evaluated using feedback obtained from the previous step, with the typical choice of feedback law being 
\begin{equation}
    \beta_j = -A_{j-1}
    \label{eq:discreteFeedback}
\end{equation}
with $A_{j-1} = i\bra{\psi_{j-1}} \, \bigl[H_{\text{d}},H_{\text{p}}\bigr]\,\ket{\psi_{j-1}}$. Based on these principles, the steps for executing FALQON are depicted in Fig. \ref{fig:FALQON protocol.} and described in the associated figure caption. 
Crucially, at each step the quantity $A_{j-1}$ is obtained by estimating the expectation value of $i[H_{\mathrm{d}},H_{\mathrm{p}}]$. This observable can, in general, be expanded into a linear combination of Pauli operators. These Pauli operators can be grouped into mutually commuting subsets, and the expectation values of each subset can be estimated separately. Then, the output results can be combined classically to evaluate $A_{j-1}$ \cite{larsen24}. Thus, the feedback law is experimentally accessible, but it can require many repeated circuit executions at each layer to estimate the required expectation value.
We refer to the FALQON sequence $\{\beta_j\}_{j=1}^{\ell}$ produced by the feedback law in Eq.~\eqref{eq:discreteFeedback} as a {parameter curve}. This is analogous to the $(\boldsymbol\gamma,\boldsymbol\beta)$ parameter sequences in QAOA and the continuous annealing schedules defined by $A(t),B(t)$ in QA. 

To generate a FALQON parameter curve, each expectation value $A_{j-1}$ is estimated from qubit measurements, performed on repeated executions of the FALQON circuit up to layer $j-1$. In practice, FALQON can require deep circuits with many layers, increasing the measurement cost.
In the following, we explore the prospect of training a machine learning model to predict these curves classically, thereby removing the need for qubit measurements.

\section{Machine Learning prediction of FALQON Parameter Curves}
\label{sec:model}

This section presents the ML model we use to predict FALQON parameter curves 
for two classes of problems: MaxCut on weighted graphs and transverse-field Ising models on graphs. The former is a widely studied combinatorial optimization problem that has been extensively studied in both classical \cite{GoemansWilliamson1995} and quantum settings, and has been utilized as a standard benchmark for QAOA and FALQON \cite{farhi14,magann22}. The latter is also a widely utilized benchmark problem, and expands the scope of our work to ``quantum'' Hamiltonians, i.e., whose ground states can possess entanglement. We first overview the application of FALQON towards each of these problem classes, and then go on to motivate and describe the high-level architecture of the ML model, with further details captured in Figure~\ref{fig:Proposed Hybrid ML Architecture.}. We then discuss the training procedure used in this work. Additional architectural details, a list of hyperparameters, and alternative pooling and readout choices are provided in Appendix~\ref{app:ml}. 

\subsection{FALQON for MaxCut}

Given an undirected graph $G=(V,E)$ with vertex set $V$ of size $|V|=n$ and edge set $E$, a {cut} is a partition of $V$ into two disjoint subsets. The cut size is the number of edges that cross the partition, and the (unweighted) MaxCut problem asks for the partition that maximizes this number. In the weighted variant we consider here, each edge is assigned a positive weight $w_{ij}$, and the objective becomes to maximize the total weight of edges crossing the cut. Figure~\ref{fig:FALQON protocol.} illustrates this premise in the case of a 3-node graph, where the dashed line partitioning the graph on the right-hand side of the figure represents the maximum cut.

The weighted MaxCut problem can be cast as a ground-state preparation task by defining the problem Hamiltonian
\begin{equation}
    H_{\mathrm p}^{\mathrm{MC}}=-\sum_{(i,j)\in E} w_{ij}\,\frac{1-Z_iZ_j}{2},
\label{eq:problemHamiltonian}
\end{equation}
whose ground state encodes the maximum cut, where $Z_i$ denotes the Pauli-$Z$ operator on qubit $i$ and $w_{ij}$ are the edge weights. The driver Hamiltonian considered in the present work is given by
\begin{equation}
    H_{\text{d}} = \sum_{i=1}^n X_i,
\label{eq:driverHamiltonian}
\end{equation}
where $X_i$ denotes the Pauli-$X$ operator on qubit $i$. Qubits are initialized in the ground state of $H_{\text{d}}$, given by
\begin{equation}
    \ket{\psi_0} \;=\; \ket{-}^{\otimes n},
\label{eq:initialstate}
\end{equation}
where $\ket{-}=\frac{1}{\sqrt{2}}(\ket{0}-\ket{1})$.

We evaluate performance using two figures of merit. The first is the {approximation ratio}, defined at layer $j$ as 
\begin{equation}
    r_{A,j}=\frac{E_j}{E_{\min}},
\label{eq:approxratio}
\end{equation}
where $E_j = \bra{\psi_j}H_{\mathrm p}^{\mathrm{MC}}\ket{\psi_j}$ and $E_{\min}$ denotes the minimum eigenvalue of the problem Hamiltonian. 
In our numerical simulations, $E_{\min}$ is computed exactly and used only to evaluate performance. It is not assumed to be known during execution of FALQON or during inference with the trained ML model. In a setting where exact normalization is unavailable, one could instead report the energy directly, compare against the best sampled cut value, or estimate approximation quality relative to a classical benchmark when available.

The second figure of merit we consider is the squared overlap with the ground-state-subspace, defined at the $j$th layer as
\begin{equation}
    \phi_j=\sum_i\bigl|\langle g_i|\psi_j\rangle\bigr|^2,
\label{eq:sucessprob}
\end{equation}
where $\{\ket{g_i}\}$ denotes the set of degenerate ground states of $H_{\mathrm p}^{\mathrm{MC}}$. Equivalently, $\phi_j$ is the fidelity with the ground-state subspace of $H_{\mathrm p}^{\mathrm{MC}}$. For MaxCut, this quantity can also be interpreted as the probability that a computational-basis measurement samples an optimal cut. Thus, $r_{A,j}$ captures the average cut quality, while $\phi_j$ captures the probability weight assigned to globally optimal solutions.

\subsection{FALQON for the transverse-field Ising model}
\label{sec:tfim}

The transverse-field Ising model (TFIM) describes interacting spins acted on by a transverse field. Here, we consider spins with pairwise interactions, where the coupling graphs are sampled from the same distributions as the graphs considered for MaxCut. The TFIM Hamiltonian is given by 
\begin{equation}
    H_{\mathrm p}^{\mathrm{TFIM}}=H_{ZZ}+hH_X,
\label{eq:tfim_hamiltonian}
\end{equation}
with
\begin{equation}
    H_{ZZ}=\sum_{(i,j)\in E}w_{ij}Z_iZ_j,
\qquad
H_X
=
\sum_{i=1}^{n}X_i .
\label{eq:tfim_terms}
\end{equation}
Here, $w_{ij}$ denotes the edge weight and $h$ is the transverse-field strength. In the numerical experiments below we set $h=1$. We sample the edge weights independently from the uniform distribution on $[0,2]$. Also, the unweighted baseline is obtained by setting all edge weights equal to one. For studies of the TFIM, we consider the same driver Hamiltonian and initial state given in Eqs. (\ref{eq:driverHamiltonian}) and (\ref{eq:initialstate}).

Since the transverse-field term in $H_{\mathrm p}^{\mathrm{TFIM}}$ is proportional to the driver Hamiltonian, one layer of FALQON is given by
\begin{equation}
    \ket{\psi_j}=e^{-i\Delta t\,(\beta_j+h)H_X}e^{-i\Delta t\,H_{ZZ}}\ket{\psi_{j-1}},
\label{eq:tfim_update}
\end{equation}
with the feedback law generated from the commutator
\begin{equation}
    \beta_j=-A_{j-1},
    \qquad
    A_{j-1}=i\bra{\psi_{j-1}}\bigl[H_X,H_{\mathrm p}^{\mathrm{TFIM}}\bigr]\ket{\psi_{j-1}} .
\label{eq:tfim_feedback}
\end{equation}
Because $[H_X,hH_X]=0$, the feedback is determined by the noncommuting Ising component $H_{ZZ}$, while the performance is evaluated with respect to the full TFIM Hamiltonian in Eq.~\eqref{eq:tfim_hamiltonian}.

We evaluate TFIM performance using the same figures of merit $r_{A,j}$ and $\phi_j$ introduced in Eqs.~\eqref{eq:approxratio} and \eqref{eq:sucessprob}, but using $H_{\mathrm p}^{\mathrm{TFIM}}$. Unlike $H_{\mathrm p}^{\mathrm{MC}}$, whose spectrum is nonpositive by construction, the TFIM Hamiltonian $H_{\mathrm p}^{\mathrm{TFIM}}$ has an indefinite spectrum. Thus, for an arbitrary state, the normalized quantity
$r_{A,j}^{\mathrm{TFIM}}$ could in principle be negative. The feedback law is designed to decrease $\langle H_{\mathrm p}^{\mathrm{TFIM}}\rangle_j$ with layer, and our simulations remain in the negative-energy regime. Therefore, along the exact FALQON trajectories and sufficiently accurate approximations to them, $r_{A,j}$ remains between 0 and 1. For other choices of initial state, this interpretation need not hold.

\subsection{Machine learning model}

Our ML model is based on a {teacher–student} framework: a higher-capacity teacher produces high-quality reference curves, and a lighter student is subsequently trained to approximate those curves \cite{hinton2015distillation}. The teacher is trained to fit FALQON curves directly, and the student is trained to mimic the teacher (a distillation loss on the whole curve), with an auxiliary loss on a set of predicted scalars $\hat s$. Details of the loss functions are given in Appendix~\ref{app:ml}. At a high level, the student and teacher models share the following flow. A weighted graph as input is first fed into a {graph neural network (GNN)}, where each node aggregates information from its neighbors and connected edges in a message-passing fashion~\cite{gilmer2017mpnn}. We use a GNN here because it is a natural fit for our graph Hamiltonian problems, i.e., the aggregation it performs depends only on adjacency and is permutation-invariant to vertex relabelings. This is relevant because it prevents spurious label memorization given that the MaxCut and TFIM objectives are invariant to vertex relabeling. This is followed by multiple convolutional layers, i.e., given by Graph Isomorphism Network with Edge Features (GINEConv) and the mask label prediction (TransformerConv) layers in the teacher and the student, respectively \cite{hu2020pretrain, shi2021unimp}, and the outputs are pooled to capture both local statistics and global structure. This is then sent through a small {readout} network that outputs the FALQON parameter curve $\hat\beta$, where the hat is used to denote that it is a model prediction. The teacher and student share this overall pattern; the teacher uses a more expressive, conditioned readout, whereas the student is streamlined. In our numerical analyses, we have found that simpler models, e.g., using dense networks only, ignoring graph structure, or only matching values, produce worse predictions. 

\begin{figure}[t]
\centering
\includegraphics[width=1.0\columnwidth]{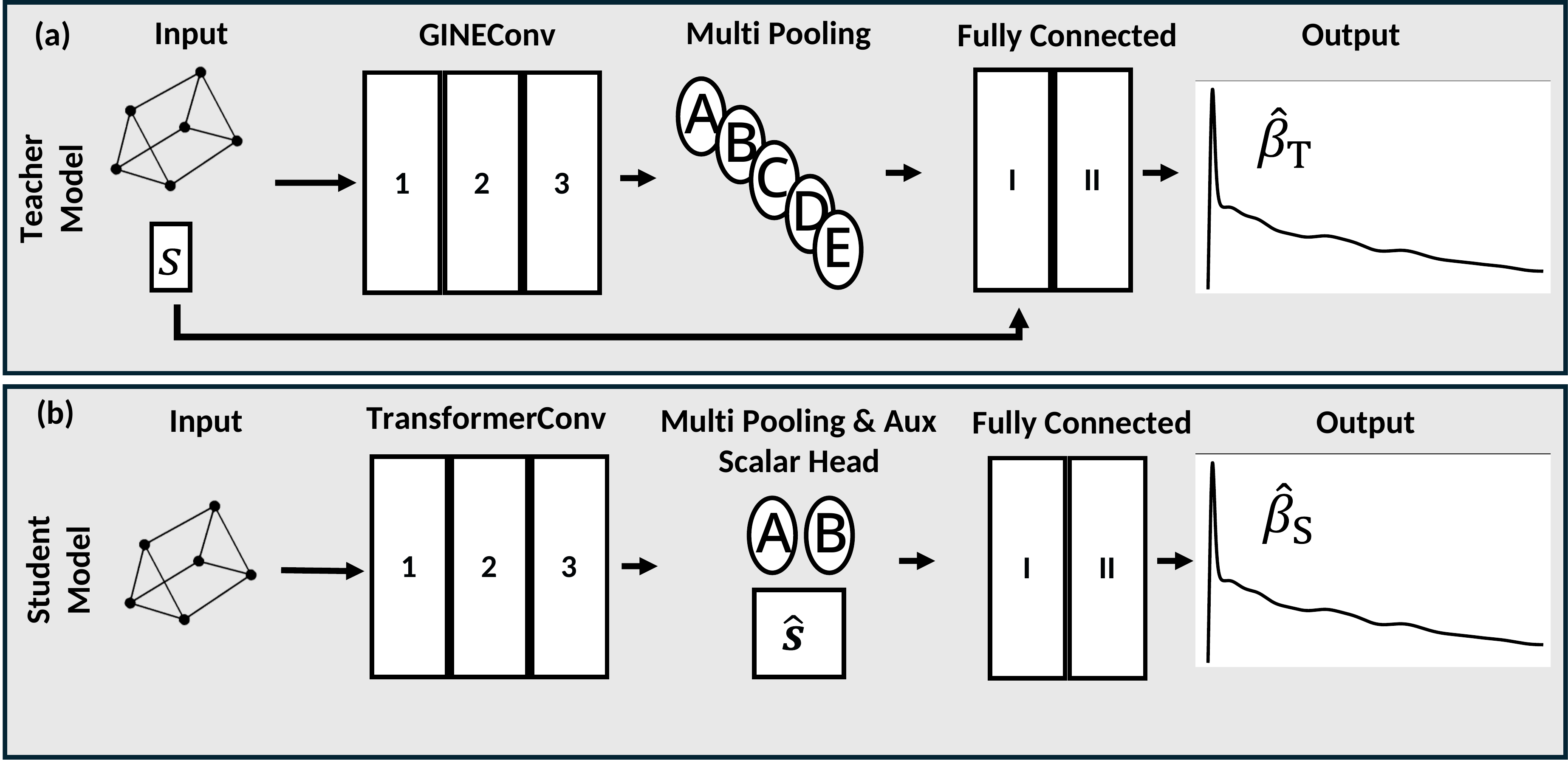}
\caption{Teacher–student ML framework for predicting FALQON parameter curves for MaxCut.
(a) Teacher: the input graph $G$ and five scalar features $s$ are processed by a graph neural network; pooled graph features feed a small conditioning module that outputs a full parameter curve $\hat\beta_{\mathrm{T}}$.
(b) Student: a lighter graph network produces an embedding $g$ and an auxiliary estimate $\hat s$ (supervised only during training); a small readout maps $[g,\hat s]$ to the predicted curve $\hat\beta_{\mathrm{S}}$.
At test time, only $\hat\beta_{\mathrm{S}}$ is used. Further details are given in Appendix~\ref{app:ml}.}

\label{fig:Proposed Hybrid ML Architecture.}
\end{figure}

We consider weighted, 3-regular graphs with $n\in\{4, 6, 8, 10, 12\}$ and with edge weights sampled uniformly from $[0, 2]$. In total we generate 2{,}240 instances and split them into $60\%$ training (of which $80\%$ are used for parameter updates and $20\%$ for validation) and $40\%$ testing. For each instance, we compute the FALQON parameter curve exactly in numerical simulation with time step $\Delta t = 0.01$. We train with batch size~1 so that each optimization step processes a full parameter curve, and we use Adam \cite{KingmaB14}. Training runs for 100 epochs with learning rate $10^{-3}$, with hyperparameters tuned using Optuna, an open source optimization toolkit that automates search over model settings \cite{optuna}. The model outputs parameter curves of length $\ell  = 1001$ to match the FALQON depth used in our numerical experiments. The forward path follows Fig.~\ref{fig:Proposed Hybrid ML Architecture.}: three message-passing blocks (1–3), multi-pooling heads (A–E), and a two-stage readout (I–II). The teacher uses a conditioned (dynamic) readout whose weights are produced by a small hypernetwork fed by the scalars $s$ described in Appendix~\ref{app:ml}; the student uses a lightweight readout and additionally predicts $\hat s$ as an auxiliary target. After training, we use only the student to predict parameter curves; the teacher is not needed at test time.

The resulting ML workflow used in this work is as follows. For the MaxCut study, we generate a training set of weighted graph instances and compute reference FALQON parameter curves by classical simulation. After training the teacher-student model, a new weighted graph is passed through the trained student model, which outputs a predicted parameter curve in a single classical inference step. This curve can then be replayed as a fixed schedule on the corresponding weighted Hamiltonian for validation, if desired.
The model architecture and workflow were designed for the MaxCut setting. The TFIM study is included as an initial exploration of whether the same learning strategy can be useful beyond MaxCut. For TFIM, the same teacher-student architecture and training procedure are used, but we train a new model on TFIM-generated FALQON curves. Thus, the TFIM results test whether the same approach and architecture can capture parameter curves for a different graph-defined Hamiltonian.
 
\section{Results}
\label{sec:results}
\begin{figure*}[t!]
  \centering
  \includegraphics[scale=0.65]{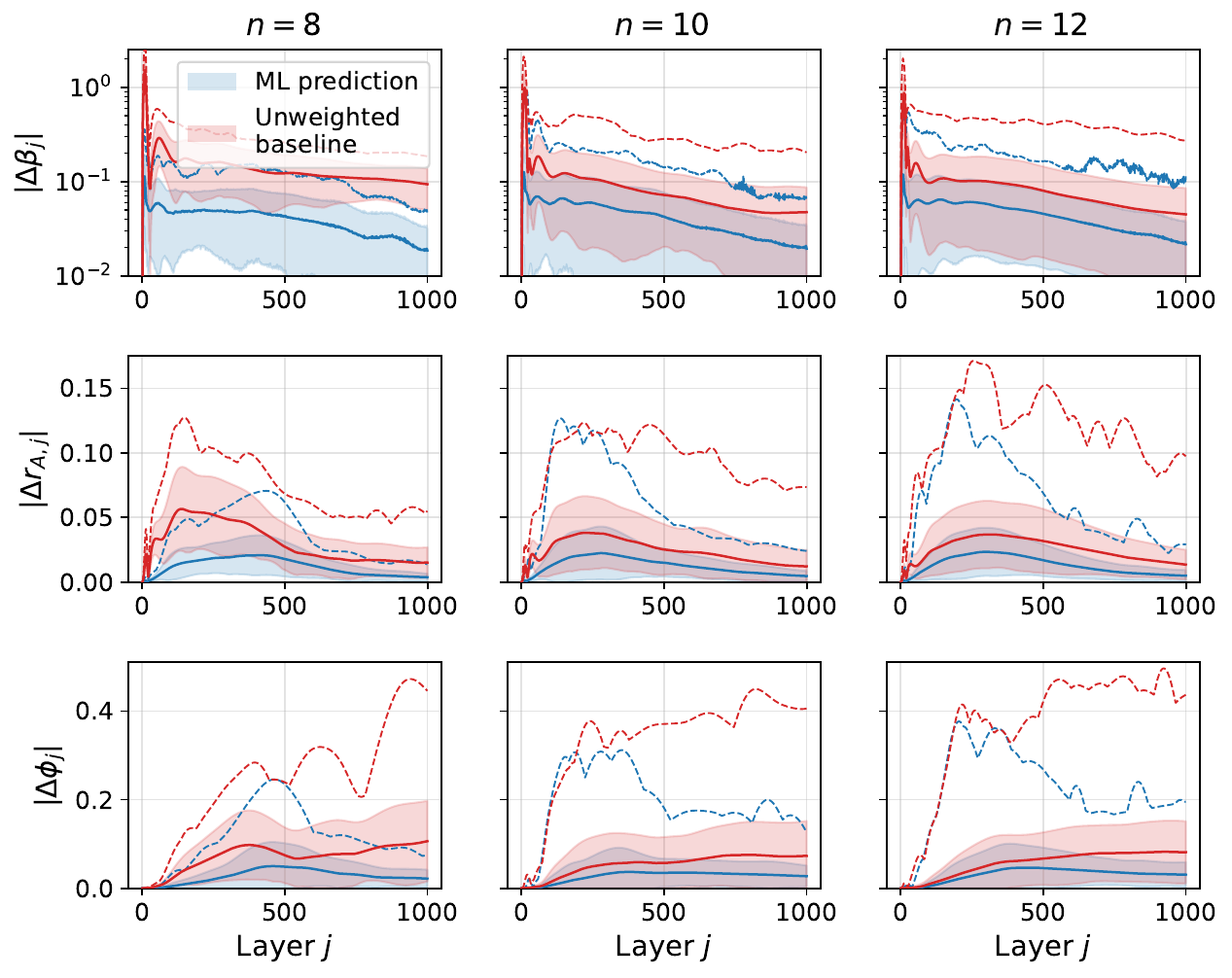}
  \caption{MaxCut performance is plotted versus layer for the ML model predictions (blue) as well as the unweighted baseline (red) for different figures of merit and different problem sizes. Solid curves show the mean across problem instances, shaded bands indicate the associated standard deviation, and dashed curves indicate the maximum error across instances. The three columns of panels correspond to graph sizes $n\in\{8,10,12\}$. The top row shows the parameter error $|\Delta\beta_j|$ on a logarithmic scale, the middle row shows the approximation-ratio error $|\Delta r_{A,j}|$, and the bottom row shows the ground-state-subspace-overlap error $|\Delta\phi_j|$. For a given weighted graph instance, these errors are computed with respect to the true FALQON curve and its performance for solving MaxCut on that instance. The ML model predictions are obtained using the same weighted graph as input, and the unweighted baseline corresponds to using the FALQON parameter curve computed for the same topology with all weights set equal to one. The logarithmic scale resolves the tail-region behavior of the parameter error, where the ML predictions remain close to the weighted FALQON reference and generally improve over the unweighted baseline.}
  \label{fig:pred_vs_unw_diff}
\end{figure*}

\begin{figure*}[t!]
  \centering
  \includegraphics[scale=0.65]{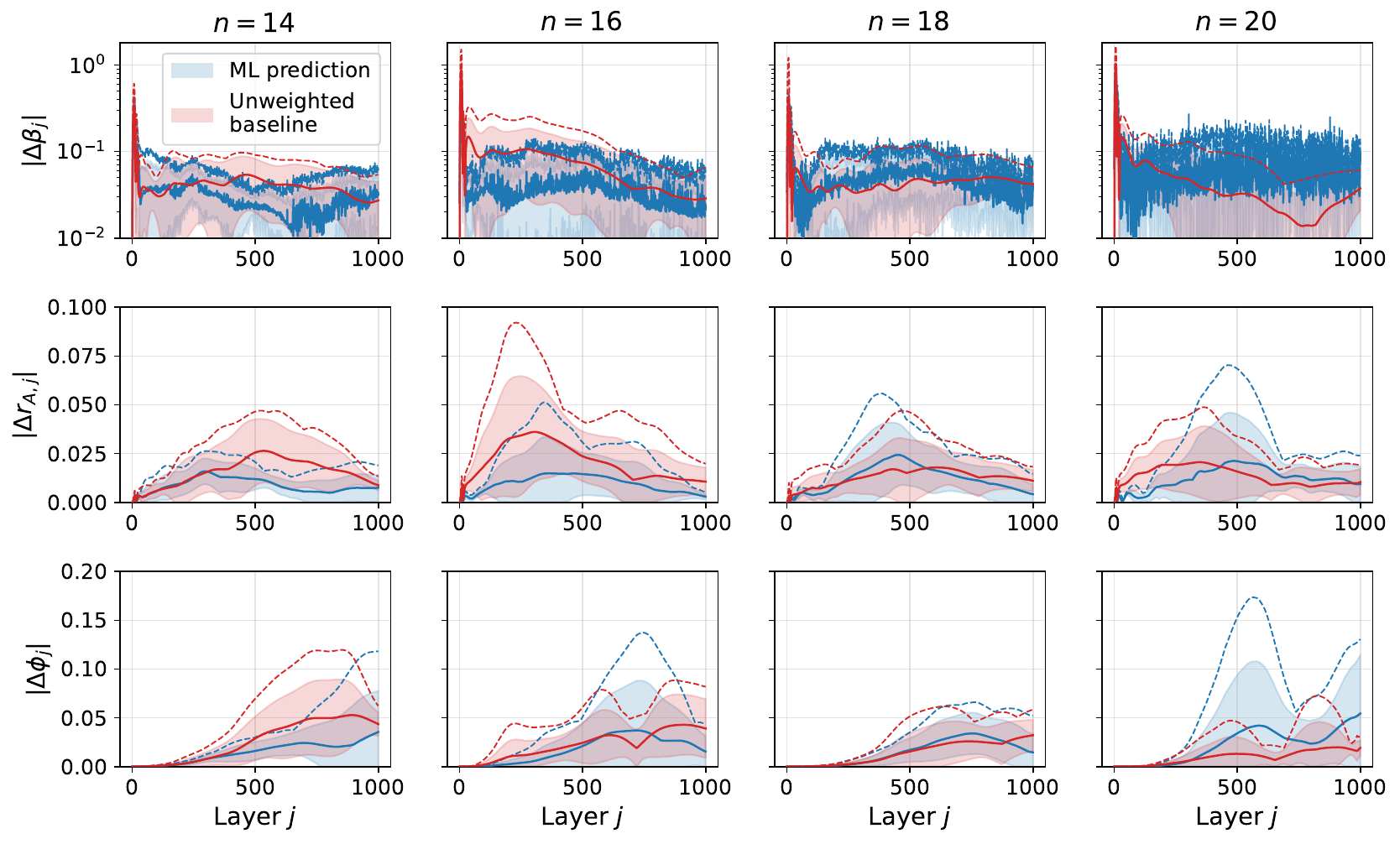}
    \caption{MaxCut performance is plotted versus layer for the ML model predictions (blue) as well as the unweighted baseline (red) for different figures of merit and for problem sizes $n\in\{14,16,18,20\}$ not seen during training. The plot conventions match those in Fig.~\ref{fig:pred_vs_unw_diff}. The logarithmic scale in the top row makes the long-layer behavior of $|\Delta\beta_j|$ visible after the initial peak region. The ML model continues to reproduce key features of the FALQON parameter curves outside the training range, although the performance gradually deteriorates as the problem size moves farther from the training distribution.}
  \label{fig:ml_falqon_abs_error_bounds}
\end{figure*}

\begin{figure}[t]
  \centering
  \includegraphics[width=0.8\linewidth]{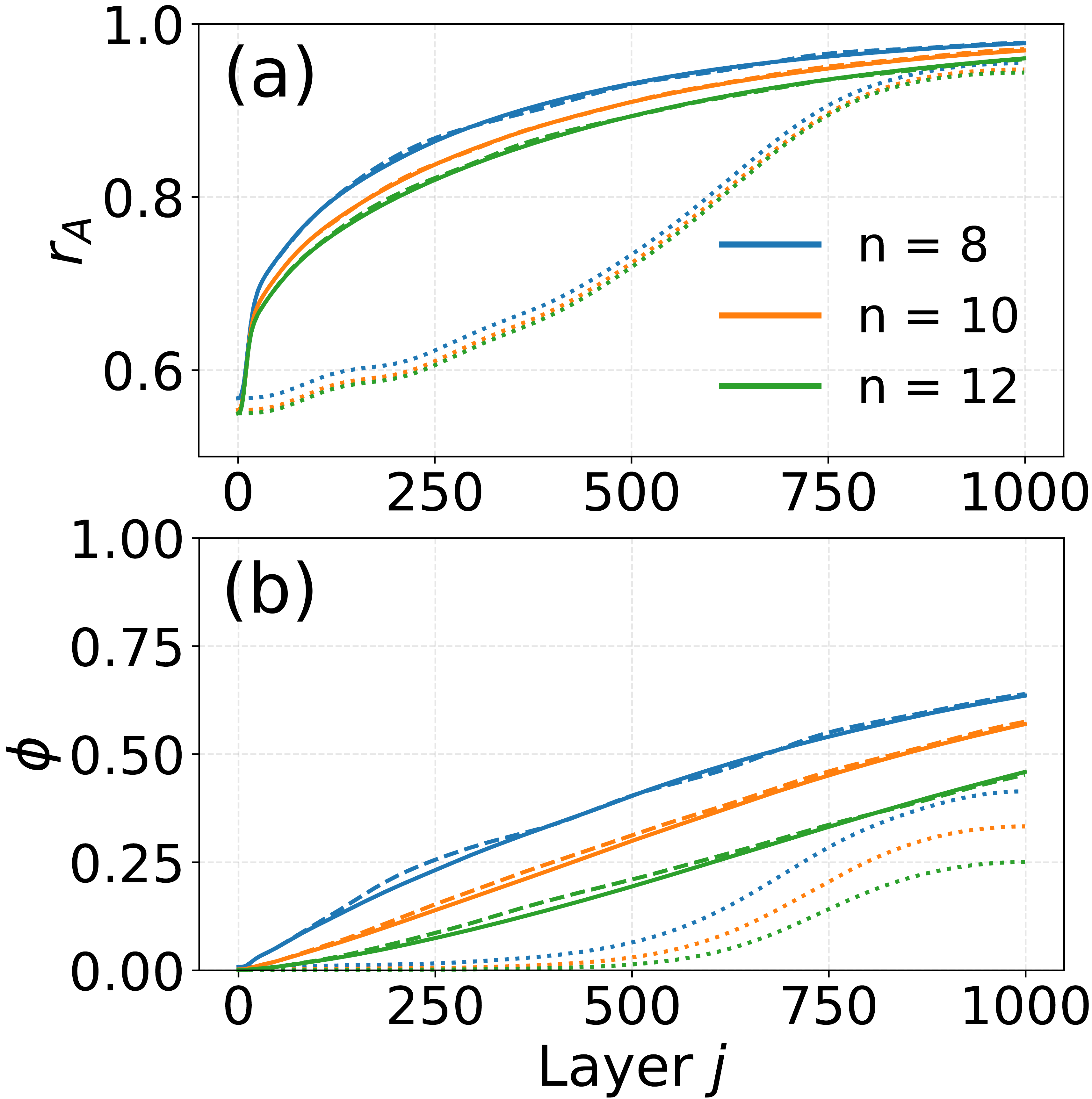}
  \caption{Mean approximation ratio and mean ground-state-subspace overlap are plotted versus layer for the exact FALQON reference (solid curves), a digitized linear-annealing baseline (dotted curves), and the ML model predictions (dashed curves) for solving weighted MaxCut. Panel (a) shows the mean approximation ratio $r_A$ and panel (b) shows the mean ground-state-subspace overlap $\phi$; for $n\in\{8,10,12\}$, the means are computed over $40,152,$ and $680$ weighted graph instances, respectively. The magnitude of the associated standard deviations (not shown) of the FALQON reference curves and ML predictions are similar, and smaller on average compared with the linear-annealing baseline. }
  \label{fig:equal-time}
\end{figure}

\begin{figure*}[t!]
  \centering
  \includegraphics[scale=0.65]{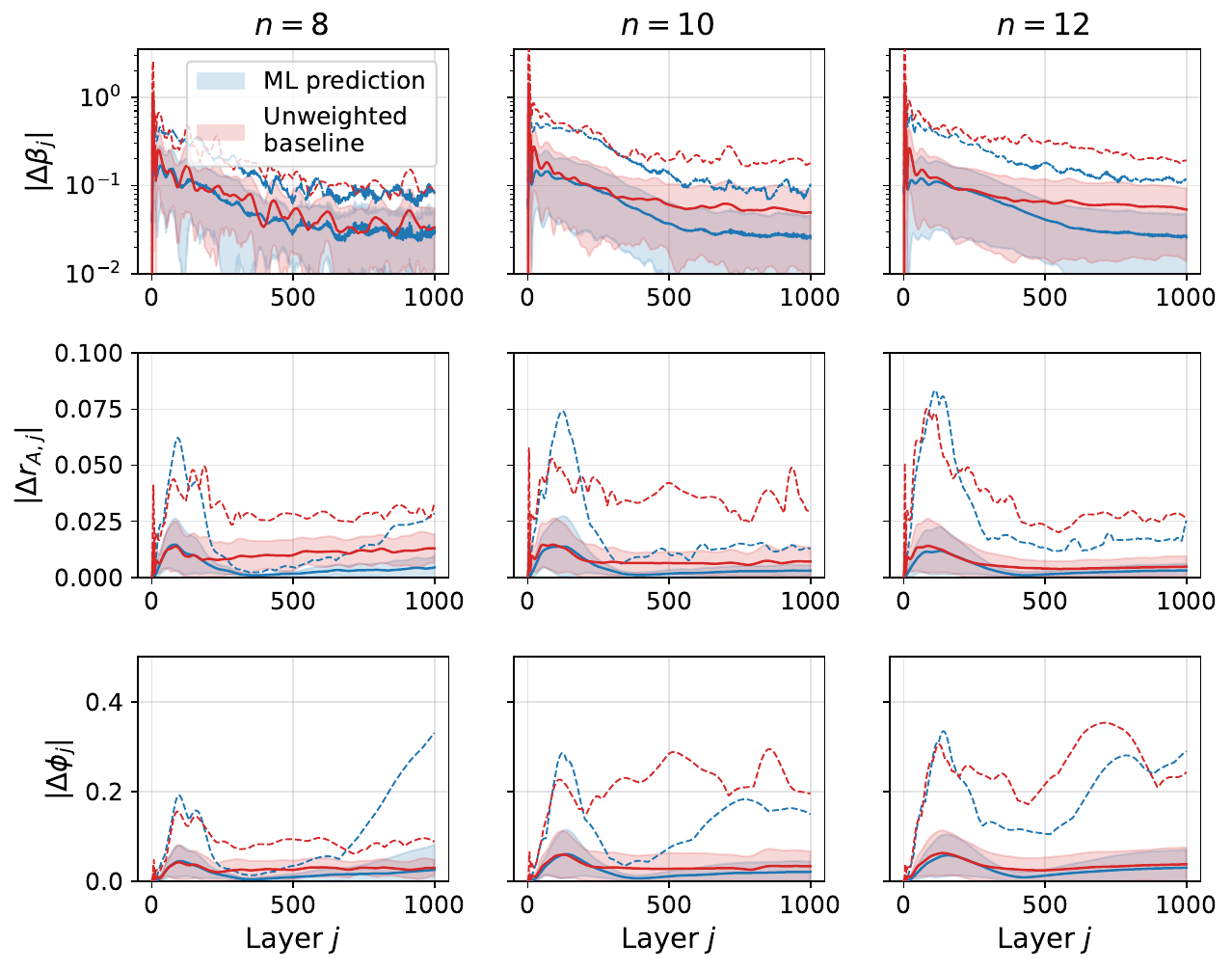}
  \caption{Performance is plotted versus layer for the TFIM test instances with $n\in\{8,10,12\}$. The ML model predictions are shown in blue and the unweighted baseline is shown in red. Solid curves show the mean across problem instances, shaded bands indicate the associated standard deviation, and dashed curves indicate the maximum error across instances. The top row shows the parameter error $|\Delta\beta_j|$ on a logarithmic scale, the middle row shows the normalized-energy error $|\Delta r_{A,j}|$, and the bottom row shows the ground-state-subspace-overlap error $|\Delta\phi_j|$. Both the ML-predicted curves and the unweighted baseline curves are replayed on the same weighted TFIM Hamiltonians and compared against the weighted FALQON reference. The ML model generally produces smaller errors than the unweighted baseline, indicating that it captures weight-dependent features of the TFIM dynamics.}
  \label{fig:tfim_seen}
\end{figure*}

\begin{figure*}[t!]
  \centering
  \includegraphics[scale=0.65]{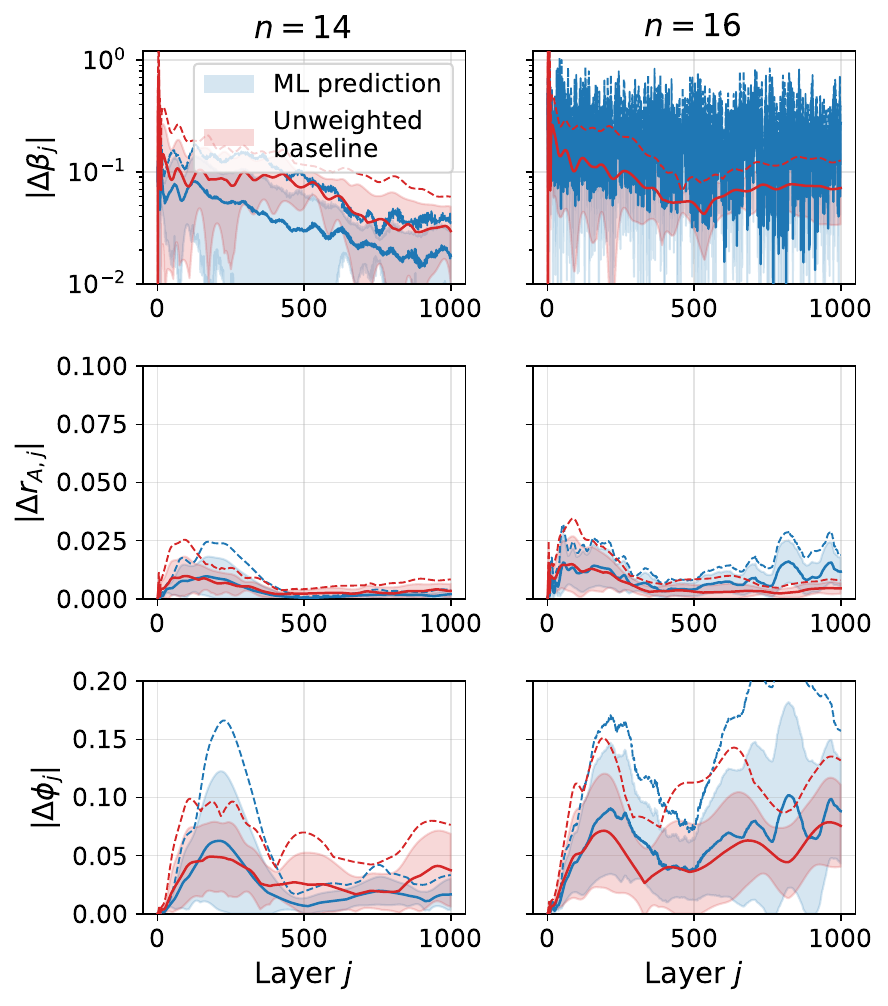}
  \caption{Extrapolation performance for TFIM instances with $n\in\{14,16\}$, which were not included in training. The plot conventions match those of Fig.~\ref{fig:tfim_seen}. For $n=14$, the ML-predicted curves remain close to the weighted FALQON reference and generally improve over the unweighted baseline. For $n=16$, the prediction becomes less stable, as shown by the larger parameter fluctuations and the increased energy and overlap errors at later layers. These results suggest that the model learns useful TFIM parameter-curve structure beyond the training sizes, while also revealing the onset of extrapolation limitations.}
  \label{fig:tfim_large}
\end{figure*}

In this section, we present numerical results exploring how the ML model described in Sec.~\ref{sec:model} performs at predicting FALQON parameter curves. We organize the section around four questions: (1) How faithfully does the student model reproduce FALQON results layer by layer for weighted MaxCut? (2) How well does the model extrapolate to MaxCut problem sizes not seen during training? (3) For MaxCut, how do the learned curves perform when compared with a standard digitized linear-annealing schedule? (4) Does the same learning strategy extend to a graph-defined transverse-field Ising model? Questions (1)--(4) are analyzed in Figs.~\ref{fig:pred_vs_unw_diff}, \ref{fig:ml_falqon_abs_error_bounds}, \ref{fig:equal-time}, and \ref{fig:tfim_seen}--\ref{fig:tfim_large}, respectively.

Figure \ref{fig:pred_vs_unw_diff} compares, layer by layer, the reference FALQON parameters $\beta_j$ with the predictions produced by our {student} ML model on weighted 3-regular graphs with $n\in\{8,10,12\}$. These results are plotted in blue. For a given weighted graph instance, the reference FALQON parameters are computed from an exact simulation of running FALQON for solving MaxCut on that weighted graph instance. The ML model predictions are obtained using the same weighted graph as input to the student model to generate an ML-predicted parameter curve. This latter ML curve is then used to parameterize an exact simulation of a FALQON circuit for solving MaxCut on this weighted graph instance, and its performance is evaluated in terms of the absolute deviations between the predicted and reference values, according to
\begin{equation}
|\Delta\beta_j| \equiv |\hat\beta_j-\beta_j|,
\label{eq:falqondev1}
\end{equation}
\begin{equation}
|\Delta r_{A,j}| \equiv \bigl|\,\hat r_{A,j}-r_{A,j}\,\bigr|,
\label{eq:falqondev2}
\end{equation}
and
\begin{equation}
|\Delta\phi_j| \equiv \bigl|\,\hat\phi_j-\phi_j\,\bigr|,
\label{eq:falqondev3}
\end{equation}
at each layer $j$, where hats denote ML student model predictions. We first observe that across $n\in\{8,10,12\}$, the ML predictions closely track FALQON for $\beta_j$. This matters because any layer-wise deviations $\Delta\beta_j$ can introduce errors that then accumulate as a function of layer. Here, we do not observe this issue, and the good agreement between the ML predictions and the FALQON reference curves extends to $r_A$ and $\phi$. 

The results in red in Fig.~\ref{fig:pred_vs_unw_diff}, meanwhile, show the performance of an ``unweighted baseline'' model, included here to benchmark the ML model results against. For a MaxCut problem instance defined by a given weighted graph, this unweighted baseline is computed as follows. First, we run FALQON for this graph, but with all edge weights set equal to 1. This outputs a parameter curve associated with solving the unweighted version of the MaxCut problem. The performance of this unweighted-baseline curve is then evaluated in simulation against the original weighted graph. For edge weights near one, this baseline is expected to be strong because the weighted Hamiltonian is close to its unweighted counterpart. As the weight distribution broadens, however, the baseline becomes a more stringent test of whether the ML model learns weight-dependent effects. By utilizing the unweighted solution as a baseline to compare against, we can therefore check our ML model's ability to learn these edge weight effects, in addition to the effects of the underlying graph topology. We see in Fig.~\ref{fig:pred_vs_unw_diff} that the ML model results are consistently better than the results of the unweighted baseline, confirming that the ML model learns {weight-dependent} parameter curves. We observe that performance is strongest at $n=12$. We attribute this to greater training diversity: there are 85 non-isomorphic graphs for $n=12$ versus 5 and 19 for $n=8$ and $n=10$, respectively; this richer variety appears to regularize the predictor and improve generalization within the training range. 

The ability of the ML model to generalize beyond the training sizes is explored next, with results presented in Fig.~\ref{fig:ml_falqon_abs_error_bounds}. For unseen graphs with
$n\in\{14,16,18,20\}$, the absolute deviations $|\Delta\beta_j|$, $|\Delta r_{A,j}|$, and $|\Delta\phi_j|$ remain small on average, with modest variability across layers. Notably, $|\Delta r_{A,j}|$ and $|\Delta\phi_j|$ stay controlled even as $n$ increases, suggesting that an ML model can serve as a reliable {warm-start} surrogate for FALQON’s iterative measurement-feedback loop on larger problem sizes.

Our final MaxCut analysis compares the learned curves with a linear digitized quantum annealing heuristic. For this comparison, we use the same number of layers $\ell$ and the same time step $\Delta t$ as in the corresponding FALQON simulation, so that the total digitized evolution time is $T=\ell\Delta t$. The linear schedule is evaluated at the discrete times $t_j=j\Delta t$, with $A(t_j)=1-t_j/T$ and $B(t_j)=t_j/T$. We do not separately optimize the total annealing time for each problem size or instance. Thus, this comparison should be interpreted as an equal-layer, equal-total-time comparison against a simple linear schedule, rather than as a comparison against an optimized annealing protocol.

In Fig.~\ref{fig:equal-time}, we compare the MaxCut performance, as quantified by $r_A$ and $\phi$, of the ML-predicted parameter curves against standard linear QA schedules, digitized for implementation on circuit-model quantum computers. Given that the identification of optimal QA schedules is known to be difficult, the aim of this comparison is to explore whether the ML model predictions could have value for improving performance over other more commonly-used QA schedule heuristics. In line with previous work \cite{magann22-2}, this comparison is carried out by fixing the total number of layers and comparing the methods' performance with this parameter held fixed. Dashed curves show the mean performance of the parameter curves predicted by the student model for different problem sizes, while dotted curves show the performance when a linear QA schedule is used. As a reference, exact FALQON results are shown as solid curves. We find that the ML model results track the FALQON mean curves closely for both $r_A$ and $\phi$ and consistently outperform the linear schedule across the problem sizes we consider, $n\in\{8,10,12\}$.

Having established this behavior for weighted MaxCut, we repeat the analysis for the TFIM introduced in Sec.~\ref{sec:tfim}. For each weighted TFIM instance, we compare the weighted FALQON reference curve against two alternatives in line with our analysis for the MaxCut results, i.e., the first being the ML-predicted curve obtained from the weighted graph input and then replayed on the same weighted TFIM Hamiltonian, and the second being an unweighted graph baseline, obtained by running FALQON on the corresponding unweighted graph and replaying that parameter curve on the weighted TFIM Hamiltonian.

Figure~\ref{fig:tfim_seen} shows the TFIM results for $n\in\{8,10,12\}$, organized in the same format as Fig. \ref{fig:pred_vs_unw_diff}. Across these system sizes, the ML-predicted curves remain close to the weighted FALQON reference and generally yield smaller errors than the unweighted baseline, indicating that the ML model is able to learn how the edge weights modify the FALQON trajectory for the TFIM.

The energy and overlap errors show the effect of the predicted curve on the actual TFIM dynamics. Since $r_{A,j}$ measures closeness to the ground-state energy and $\phi_j$ measures probability weight in the ground-state subspace, smaller errors indicate that the model captures weight-dependent information relevant to the ground-state-preparation behavior of the FALQON trajectory.

We next test TFIM extrapolation beyond the training sizes, with results for $n\in\{14,16\}$ shown in Fig.~\ref{fig:tfim_large}. For $n=14$, the ML predictions remain reasonably close to the weighted FALQON reference and generally improve over the unweighted baseline. This behavior is consistent with the MaxCut extrapolation results and suggests that the model has learned structure that is useful beyond the exact training sizes. For $n=16$, however, the quality of the extrapolation degrades, suggesting a finite generalization window. That is, the trained ML model captures useful low-dimensional structure in the FALQON parameter curves, but its reliability decreases as the system size moves farther from the training distribution. Thus, the TFIM analysis supports the broader applicability of the approach beyond MaxCut, while also showing that robust extrapolation to larger systems will likely require larger training sets or transfer-learning strategies.

\section{Conclusion and outlook}
\label{sec:outlook}

In this work, we have explored the prospect of predicting FALQON parameter curves with classical machine learning. We have found that the teacher-student ML model developed here can effectively predict FALQON parameter curves associated with MaxCut and TFIM problems when fed a weighted graph as input. For both settings, the ML model reproduces key features of the reference FALQON curves and generally improves over an unweighted baseline within the regimes studied. These results indicate that once trained, an ML-driven approach can be a viable, measurement-free, and classical-optimization-free alternative to the iterative, measurement-based feedback procedure utilized in FALQON. We have also illustrated that the ML-predicted parameter curves can offer benefits in a digitized QA setting when used in place of conventional, linear QA schedules. The latter finding opens prospects for utilizing ML to design schedules for use in adiabatic quantum state preparation more broadly. 

There are also important practical considerations for hardware implementations. First, the circuit depths considered in our simulations may be challenging for near-term quantum devices, where noise can accumulate over many layers. Second, the measurement savings discussed here concern the measurements needed for the generation of the FALQON parameter curve, i.e., as the trained ML model can remove the need for these measurements by outputting the full curve classically. This does not, however, remove the need to measure the final circuit output. In practice, one would still need to sample the terminal state to read out the desired results (e.g., problem Hamiltonian expectation value or sampled bit string).

It is also useful to relate these learned curves to ideas from quantum control and counterdiabatic driving \cite{gueryodelin2019shortcuts, sels2017minimizing, kolodrubetz2017geometry}. Counterdiabatic protocols aim to suppress diabatic transitions by adding control terms constructed from spectral information about the instantaneous Hamiltonian. FALQON curves considered here also differ qualitatively from Floquet-engineered implementations of counterdiabatic driving, in which periodic modulations of the available controls are used to synthesize approximate counterdiabatic terms \cite{claeys2019floquet}. In contrast, FALQON parameter curves studied here are dominated by a smooth peak-and-tail structure and do not exhibit a pronounced high-frequency component. The learned FALQON curves studied here are not counterdiabatic ansatzes in this formal sense, since they do not introduce additional Hamiltonian terms and are not derived from the adiabatic gauge potential. Nevertheless, they can play a related heuristic role since the learned schedule can deviate from a simple linear annealing path in a way that improves performance.

There are a variety of ways in which this work could be extended. One important aspect is to further analyze the scalability of the approach. We carried out an initial investigation of this point in Fig.~\ref{fig:ml_falqon_abs_error_bounds}, where we found that the ML predictions begin to lag for $n=18,20$, indicating limited generalization and scalability to problem sizes as they move farther beyond the largest training size of $n=12$. The TFIM results in Fig.~\ref{fig:tfim_large} exhibit a similar trend where the ML predictions remain reasonably close to the FALQON reference for $n=14$, while the quality of the extrapolation degrades for $n=16$. Looking ahead, it would be interesting to train and validate the ML model on larger problem sizes to probe its ability to handle the growing complexity that arises as $n$ is increased further. It would also be valuable to study the number of problem instances needed for training ML models, and how this scales as the models are extended. In the present work, we utilized 1,344 instances for training our model, where the generation of each instance involved an exact numerical simulation of FALQON to output the reference parameter curve. Given that exact numerical simulations have costs that scale exponentially, in general, with the size of the problem, training data could alternatively be obtained from runs of FALQON on quantum hardware. Either way, the cost of generating sufficient training data must be balanced against the utility and scalability of the resulting ML model. Another direction that could support scalability would be to explore training an ML model to learn a small set of parameters entering into a parameterized function defining a FALQON parameter curve, rather than the full curve directly. The use of a parameterized function could constitute a more parsimonious approach with the potential to scale more easily, if a suitably expressive functional form could be identified. 

Another future direction could involve extending our approach beyond the problem classes considered here, by exploring how ML models trained on one ensemble of graphs (e.g., defined by fixed degree, given edge-weight distribution, or topology) perform when applied to different graph families or weight priors. Techniques such as domain adaptation \cite{domain1, domain2, domain3} or meta-learning \cite{meta1, meta2} could be explored for helping ML models adjust to unseen instances, preserving performance without retraining from scratch.  Broadening out further, ML models could also be explored for transfer between different problem classes, such as MaxCut and TFIM, as well as to additional problems to test the flexibility of the method and reveal the specific features of the problem that a generalized architecture must capture. In the present work, separate models are trained for MaxCut and TFIM, so whether a model trained on one Hamiltonian family can transfer directly to another remains an open question. Yet another aspect that could be considered in future work is to explore the impact of noise on the predictions, given that today's quantum computers exhibit numerous types of noise including pulse distortions, calibration errors, and decoherence. Given the freedom in choosing the driver Hamiltonian, another possibility could be to train an ML model to select it, or to output a targeted, layer-wise sequence of driver Hamiltonians to use. Across these different future directions, it will be important to identify suitable points of comparison to benchmark ML models' performance. 
Finally, the results reported here also open the prospect of using ML models, like the one we have designed, for predicting quantum controls, which could have broad applications across quantum science and technology. This prospect is in line with recent work using ML to learn {control} fields directly from classical optimal-control solvers \cite{khaleghian2025developmentneuralnetworkbasedoptimal}.

\begin{acknowledgments}

We gratefully acknowledge discussions with Jennie Si, James Larsen, Dominic Messina, Prakriti Biswas, and Lili Ye. This research was supported in part by an appointment to the National Nuclear Security Administration Minority Serving Institutions Internship Program (NNSA-MSIIP), sponsored by the National Nuclear Security Administration and administered by the Oak Ridge Institute for Science and Education. This work was supported by the Laboratory Directed Research and Development program (Projects $225952$ and $233972$) at Sandia National Laboratories, a multimission laboratory managed and operated by National Technology and Engineering Solutions of Sandia LLC, a wholly owned subsidiary of Honeywell International Inc. for the U.S. Department of Energy’s National Nuclear Security Administration under contract DE-NA$0003525$. This paper describes objective technical results and analysis. Any subjective views or opinions that might be expressed in the paper do not necessarily represent the views of the U.S. Department of Energy or the United States Government. SAND2026-15682O.
\end{acknowledgments}
\section*{Data Availability}

The data that support the findings of this article are not
publicly available. The data are available from the authors
upon reasonable request.

\vspace{12pt}

\appendix
\section{Model Architecture and Training Details}
\label{app:ml}

In this appendix, we provide further details on the ML model architecture and training procedure used in this work.

\begin{table*}[t]
\centering
\caption{Model hyperparameters (Optuna tuned unless marked \emph{fixed}).}
\label{tab:hyperparams}

\begin{tabular}{l l}
\hline
\textbf{Hyperparameter} & \textbf{Value} \\
\hline
\multicolumn{2}{l}{\textbf{Teacher (\emph{fixed})}}\\
Hidden dimension & 96 \\
Hypernetwork hidden & 224 \\
Output length $\ell$ & 1001 \\
GlobalAttention / Set2Set & True / True \\
\hline
\multicolumn{2}{l}{
\textbf{Nominal Student [TransformerConv; loss weights in Eq.~\eqref{eq:Ldistill}]}
}\\
$c_{1}$ (level match) & 1.9700633 \\
$c_{2}$ (slope match) & 0.3553681 \\
$c_{3}$ (curvature match) & 1.7610852 \\
$c_{4}$ (spectral penalty) & $8.7863\times 10^{-2}$ \\
$c_{5}$ (total variation) & $5.5548618\times 10^{-1}$ \\
$c_{6}$ (auxiliary scalar head) & tuned; best $c_{6}^{\star}$ \\
\hline
\end{tabular}

\vspace{2pt}
\footnotesize
\emph{Tunable values selected via Optuna (TPE) on validation loss; teacher settings are fixed.}
\end{table*}

\subsection{Data, features, scalars, and normalization}

Each MaxCut instance is a weighted 3-regular graph $G=(V,E,w)$ with positive edge weights $w_{ij}$. We encode each graph into the input tensors of a Graph Neural Network (GNN) encoder, i.e., a matrix that contains information about node features and a list with the edge attributes. Because the graphs do not have intrinsic node attributes, we use a constant node feature $x_i=1$ for every vertex $i$; all information enters the GNN through connectivity and edge weights. In addition to message passing on the graph (see Sec.~\ref{secA2}), during training both teacher and student receive a small set of scalar features $s$ that contains: the node count $|V|$; the ground state eigenvalue of the problem Hamiltonian; the spectral gap of the problem Hamiltonian; the spectral gap of the midpoint Hamiltonian, defined as $(H_{\text{d}}+H_{\text{p}})/2$; and an optional index feature that is determined by the index (after canonical labeling) of the minimum diagonal entry of $H_{\text{p}}$, where canonical labeling means a deterministic relabeling that assigns the same vertex ordering to isomorphic graphs so that indices are comparable across instances. We found the index feature to have negligible effect in our tests, so we omit it from the nominal model. To provide explicit size cues, we append $\sqrt{|V|}$ and $\log |V|$ to the set of scalar features. Thus, we have five instance scalars (if the optional index feature is included) plus two size cues. All scalar components are standardized using training-split statistics (zero mean, unit variance), and the same procedure is applied at validation and test time. We train with batch size 1.

\subsection{Encoder (message passing)}\label{secA2}

Given a weighted graph, a GNN encoder performs message passing for $\ell_{\mathrm{mp}}$ GNN layers. Here, message passing refers to each node repeatedly gathering information from its neighboring nodes, which are weighted by the edges connecting them, and updating its own internal vector to reflect that neighborhood. That is, in each layer, every node updates its embedding by aggregating a learned function of its neighbors’ embeddings and the incident edge weights. After $\ell_{\mathrm{mp}}$ layers, the encoder produces node embeddings $\{h_v^{(\ell_{\mathrm{mp}})}\in\mathbb{R}^d: v\in V\}$, i.e., learned $d$-dimensional vectors that summarize each node’s local weighted neighborhood. Because the neighbor-aggregation step is permutation invariant, the encoder is invariant to vertex relabelings.

\subsection{Pooling and the graph-level vector}

After the neighbor-update steps, each node $v$ carries a vector $h_v^{(\ell_{\mathrm{mp}})}$ that encodes its local environment. To describe the whole graph with a single object, we combine these node vectors into one fixed-length graph-level vector $z$. We do this by taking a few simple summaries over all nodes,
\[
\mu=\tfrac{1}{|V|}\sum_{v}h_v,\qquad
\sigma=\sum_{v}h_v ,\qquad
\pi=\max_{v}h_v ,
\]
and by using two standard learned summaries that adaptively weight or scan over the set of nodes: \emph{GlobalAttention} (a learned, weighted average) and \emph{Set2Set} (a small recurrent module applied to the unordered node set)\,%
\cite{li2017gatedgraphsequenceneural, vinyals2016order, gilmer2017mpnn}.
We then concatenate these pieces to form
\[
z=[\,\mu;\sigma;\pi;g_{\mathrm{GA}};g_{\mathrm{S2S}}\,],
\]
such that $z$ collects both simple whole-graph statistics (mean/sum/max) and learned global summaries that can capture more structured correlations across the graph.

\subsection{Readouts and teacher–student roles}

\paragraph{Teacher (conditioned readout).}
The teacher stacks three message-passing convolutional blocks (GINEConv with edge attributes) to produce node embeddings, applies the multi-pooling heads A-E (A: mean, B: sum, C: max, D: GlobalAttention, E: Set2Set), and concatenates them to $z$. A small \emph{hypernetwork} $H$ then takes five graph-level scalars $s$ and outputs the weights and bias $(W(s),b(s))$ of a two-stage readout (I-II), yielding the reference curve
\[
\hat\beta_{\mathrm{T}}=W(s)\,z+b(s)
\]
This lets the final map adapt smoothly across instance sizes and coarse difficulty without retraining upstream layers.

\paragraph{Student (lightweight backbone with auxiliary scalar head).}
The student uses three \emph{TransformerConv} layers with edge attributes, the same multi-pooling to obtain $z$, and two heads: (i) an auxiliary scalar head $f_{\mathrm{scal}}$ that predicts $\hat s$ (supervised only during training), and (ii) a compact two-layer Multi Layer Perceptron (MLP) (\emph{curve head}) that takes $[\,z,\hat s\,]$ and outputs $\hat\beta^\mathrm{S}$. Having the student also learn the graph statistics $s$ encourages it to capture the same information that guides the teacher. At inference the student builds a graph summary $z$, forms an internal estimate $\hat s$ from this summary using a small readout trained to match $s$, and then uses $[z,\hat s]$ to output the curve $\hat\beta^{\mathrm{S}}$; we report only $\hat\beta^{\mathrm{S}}$ and do not use the teacher.

\subsection{Distillation objective (values, differences, and light smoothing)}

By “distill” we mean that the student matches the teacher {as a function of layer index}, minimizing discrepancy on the full curve and its first/second finite differences, while lightly suppressing high-frequency oscillations. Let $\nabla$ and $\nabla^2$ denote first and second finite differences along the layer axis, $\mathcal{F}$ the real-input FFT, and $\mathrm{TV}$ total variation. The training objective is
\begin{align}
L_{\text{distill}}
&= c_{1}\,\|\hat\beta^{S}-\hat\beta^{T}\|_2^2
 + c_{2}\,\|\nabla \hat\beta^{S}-\nabla \hat\beta^{T}\|_2^2 \nonumber\\
&\quad + c_{3}\,\|\nabla^2 \hat\beta^{S}-\nabla^2 \hat\beta^{T}\|_2^2
 + c_{4}\,\sum_{\omega}\omega^{4}\,|\mathcal{F}\{\hat\beta^{S}\}(\omega)|^2 \nonumber\\
&\quad + c_{5}\,\mathrm{TV}(\hat\beta^{S})
 + c_{6}\,\|\hat s - s\|_2^2.
\label{eq:Ldistill}
\end{align}
\noindent\textit{Definitions (finite differences and transforms).}
\begin{align}
(\nabla \beta)_j &= \beta_{j+1}-\beta_j,\qquad 0 \le j \le \ell-2, \\
(\nabla^2 \beta)_j &= \beta_{j+1}-2\beta_j + \beta_{j-1},\quad 1 \le j \le \ell-2, \\
\mathcal{F}\{\beta\}(\omega_m) &= \sum_{j=0}^{\ell-1} \beta_j\, e^{-i\,\omega_m j},
\ \omega_m \coloneqq \tfrac{2\pi m}{\ell},\ m=0,\dots,\ell-1, \\
\mathrm{TV}(\beta) &\coloneqq \sum_{j=0}^{\ell-2} \lvert \beta_{j+1}-\beta_j \rvert.
\end{align}
\noindent\textit{Intuition.} $c_1$ matches level; $c_2$ matches slope (peak placement and tail decay rate); $c_3$ matches curvature (peak sharpness and tail bend); $c_4$ reduces high-frequency chatter (via $\omega^4$); $c_5$ discourages step-to-step jumps; $c_6$ aligns the scalar conditioning channel. Values given in Table \ref{tab:hyperparams}.

\subsection{Data, training protocol, and hyper-parameters}

We consider 3-regular graphs with $n\in\{4,6,8,10,12\}$. We generate 2{,}240 instances and split them into $60\%$ training (of which $80\%$ are used for parameter updates and $20\%$ for validation) and $40\%$ testing. We train for 100 epochs with Adam (learning rate $10^{-3}$), batch size~1 (per-graph curve supervision). The model outputs curves of length $\ell=1001$ to match the FALQON depth used in our numerical experiments. The coefficients in Eq.~\eqref{eq:Ldistill} are those in Table~\ref{tab:hyperparams}. The forward path follows Fig.~\ref{fig:Proposed Hybrid ML Architecture.}: three message-passing blocks (1-3), multi-pooling heads (A-E), and a two-stage readout (I-II). The teacher uses a conditioned (hypernetwork) readout fed by $s$; the student adds the auxiliary scalar head $\hat s$ which is estimated during training by a small MLP and included in the loss function. Training the nominal student on a single GPU takes on the order of hours, while a trained student predicts a full curve in milliseconds per graph.

\bibliography{BibliographyLatest}

@misc{farhi14,
  title         = {A Quantum Approximate Optimization Algorithm},
  author        = {Farhi, Edward and Goldstone, Jeffrey and Gutmann, Sam},
  year          = {2014},
  eprint        = {1411.4028},
  archivePrefix = {arXiv},
  primaryClass  = {quant-ph},
  journal       = {arXiv [quant-ph]},
}

@article{cerezo21,
  author  = {Cerezo, M. and Arrasmith, A. and Babbush, R. and others},
  title   = {Variational Quantum Algorithms},
  journal = {Nat. Rev. Phys.},
  year    = {2021},
  volume  = {3},
  pages   = {625--644},
  doi     = {10.1038/s42254-021-00348-9}
}

@article{magann22,
  author  = {Magann, Alicia B. and Rudinger, Kenneth M. and Grace, Matthew D. and Sarovar, Mohan},
  title   = {Feedback-Based Quantum Optimization},
  journal = {Phys. Rev. Lett.},
  volume  = {129},
  pages   = {250502},
  year    = {2022},
  doi     = {10.1103/PhysRevLett.129.250502}
}

@article{magann22-2,
  author  = {Magann, A. B. and Rudinger, K. M. and Grace, M. D. and Sarovar, M.},
  title   = {Lyapunov-control-inspired strategies for quantum combinatorial optimization},
  journal = {Phys. Rev. A},
  volume  = {106},
  pages   = {062414},
  year    = {2022},
  month   = dec,
  doi     = {10.1103/PhysRevA.106.062414}
}

@article{larsen24,
  author  = {Larsen, J. B. and Grace, M. D. and Baczewski, A. D. and Magann, A. B.},
  title   = {Feedback-based quantum algorithms for ground state preparation},
  journal = {Phys. Rev. Res.},
  volume  = {6},
  number  = {3},
  pages   = {033336},
  year    = {2024},
  month   = sep,
  doi     = {10.1103/PhysRevResearch.6.033336}
}

@article{lucas14,
  author  = {Lucas, Andrew},
  title   = {Ising Formulations of Many NP Problems},
  journal = {Front. Phys.},
  volume  = {2},
  pages   = {5},
  year    = {2014},
  month   = feb,
  doi     = {10.3389/fphy.2014.00005}
}

@article{PhysRevX.7.021027,
  title = {Optimizing Variational Quantum Algorithms Using Pontryagin's Minimum Principle},
  author = {Yang, Zhi-Cheng and Rahmani, Armin and Shabani, Alireza and Neven, Hartmut and Chamon, Claudio},
  journal = {Phys. Rev. X},
  volume = {7},
  issue = {2},
  pages = {021027},
  numpages = {10},
  year = {2017},
  month = {May},
  publisher = {American Physical Society},
  doi = {10.1103/PhysRevX.7.021027},
  url = {https://link.aps.org/doi/10.1103/PhysRevX.7.021027}
}

@misc{brandao2018fixedcontrolparametersquantum,
      title={For Fixed Control Parameters the Quantum Approximate Optimization Algorithm's Objective Function Value Concentrates for Typical Instances}, 
      author={Fernando G. S. L. Brandao and Michael Broughton and Edward Farhi and Sam Gutmann and Hartmut Neven},
      year={2018},
      eprint={1812.04170},
      archivePrefix={arXiv},
      primaryClass={quant-ph},
      url={https://arxiv.org/abs/1812.04170}, 
}

@article{PhysRevResearch.4.033029,
  title = {Adaptive quantum approximate optimization algorithm for solving combinatorial problems on a quantum computer},
  author = {Zhu, Linghua and Tang, Ho Lun and Barron, George S. and Calderon-Vargas, F. A. and Mayhall, Nicholas J. and Barnes, Edwin and Economou, Sophia E.},
  journal = {Phys. Rev. Res.},
  volume = {4},
  issue = {3},
  pages = {033029},
  numpages = {9},
  year = {2022},
  month = {Jul},
  publisher = {American Physical Society},
  doi = {10.1103/PhysRevResearch.4.033029},
  url = {https://link.aps.org/doi/10.1103/PhysRevResearch.4.033029}
}

@misc{tyagin2025qaoagptefficientgenerationadaptive,
      title={QAOA-GPT: Efficient Generation of Adaptive and Regular Quantum Approximate Optimization Algorithm Circuits}, 
      author={Ilya Tyagin and Marwa H. Farag and Kyle Sherbert and Karunya Shirali and Yuri Alexeev and Ilya Safro},
      year={2025},
      eprint={2504.16350},
      archivePrefix={arXiv},
      primaryClass={quant-ph},
      journal= {arXiv [quant-ph]}, 
}

@article{doi:10.1137/S0097539704445226,
author = {Kempe, Julia and Kitaev, Alexei and Regev, Oded},
title = {The Complexity of the Local Hamiltonian Problem},
journal = {SIAM J. Comput.},
volume = {35},
number = {5},
pages = {1070-1097},
year = {2006},
doi = {10.1137/S0097539704445226},

URL = { 
    
        https://doi.org/10.1137/S0097539704445226
    
    

}
}

@misc{farhi00,
  title         = {Quantum Computation by Adiabatic Evolution},
  author        = {Farhi, E. and Goldstone, J. and Gutmann, S. and Sipser, M.},
  year          = {2000},
  eprint        = {quant-ph/0001106},
  archivePrefix = {arXiv},
  primaryClass  = {quant-ph},
  journal       = {arXiv [quant-ph]},
}

@article{kadowaki98,
  author  = {Kadowaki, T. and Nishimori, H.},
  title   = {Quantum Annealing in the Transverse Ising Model},
  journal = {Phys. Rev. E},
  volume  = {58},
  pages   = {5355--5363},
  year    = {1998},
  doi     = {10.1103/PhysRevE.58.5355}
}

@article{hadfield17,
  author  = {Hadfield, Stuart and Wang, Zhihui and O'Gorman, Bryan and Rieffel, Eleanor G. and Venturelli, Davide and Biswas, Rupak},
  title   = {From the Quantum Approximate Optimization Algorithm to a Quantum Alternating Operator Ansatz},
  journal = {Algorithms},
  volume  = {12},
  number  = {2},
  pages   = {34},
  year    = {2019},
  doi     = {10.3390/a12020034}
}

@article{xie23,
  author  = {Xie, Ningyi and Lee, Xinwei and Cai, Dongsheng and Saito, Yoshiyuki and Asai, Nobuyoshi},
  title   = {Quantum Approximate Optimization Algorithm Parameter Prediction Using a Convolutional Neural Network},
  journal = {J. Phys.: Conf. Ser.},
  volume  = {2595},
  number  = {1},
  pages   = {012001},
  year    = {2023},
  doi     = {10.1088/1742-6596/2595/1/012001}
}

@inproceedings{alam20,
  author    = {Alam, Mahabubul and others},
  title     = {Accelerating Quantum Approximate Optimization Algorithm using Machine Learning},
  booktitle = {2020 Design, Automation \& Test in Europe Conference \& Exhibition (DATE)},
  year      = {2020},
  pages     = {1--6},
  doi       = {10.23919/DATE48585.2020.9116348}
}

@article{jain22,
  author  = {Jain, Nishant and Coyle, Brian and Kashefi, Elham and Kumar, Niraj},
  title   = {Graph neural network initialisation of quantum approximate optimisation},
  journal = {Quantum},
  volume  = {6},
  pages   = {861},
  year    = {2022},
  doi     = {10.22331/q-2022-11-17-861}
}

@inproceedings{paolo24,
  author    = {Zentilini, P. and Corli, S. and Prati, E. and others},
  title     = {Emulating QAOA via Graph Neural Networks},
  booktitle = {2024 IEEE International Conference on Quantum Computing and Engineering (QCE)},
pages = {472-473},
  year      = {2024},
  doi={10.1109/QCE60285.2024.10361}
}

@inproceedings{optuna,
  author    = {Akiba, Takuya and Sano, Shotaro and Yanase, Toshihiko and Ohta, Takeru and Koyama, Masanori},
  title     = {Optuna: A Next-generation Hyperparameter Optimization Framework},
  booktitle = {Proceedings of the 25th ACM SIGKDD International Conference on Knowledge Discovery and Data Mining},
  year      = {2019},
  pages     = {2623--2631},
  doi       = {10.1145/3292500.3330701}
}

@misc{mbeng19,
  title         = {Quantum Annealing: A Journey through Digitalization, Control, and Hybrid Quantum Variational Schemes},
  author        = {Mbeng, G. B. and Fazio, R. and Santoro, G. E.},
  year          = {2019},
  eprint        = {1906.08948},
  archivePrefix = {arXiv},
  primaryClass  = {quant-ph},
  journal       = {arXiv [quant-ph]},
}

@article{sack21,
  author  = {Sack, S. H. and Serbyn, M.},
  title   = {Quantum Annealing Initialization of the Quantum Approximate Optimization Algorithm},
  journal = {Quantum},
  volume  = {5},
  pages   = {491},
  year    = {2021},
  doi     = {10.22331/q-2021-07-01-491}
}

@article{PhysRevLett.131.060602,
  title = {Self-Healing of Trotter Error in Digital Adiabatic State Preparation},
  author = {Kovalsky, Lucas K. and Calderon-Vargas, Fernando A. and Grace, Matthew D. and Magann, Alicia B. and Larsen, James B. and Baczewski, Andrew D. and Sarovar, Mohan},
  journal = {Phys. Rev. Lett.},
  volume = {131},
  issue = {6},
  pages = {060602},
  numpages = {6},
  year = {2023},
  month = {Aug},
  publisher = {American Physical Society},
  doi = {10.1103/PhysRevLett.131.060602},
  url = {https://link.aps.org/doi/10.1103/PhysRevLett.131.060602}
}

@article{PhysRevLett.102.130503,
  title = {Preparing Ground States of Quantum Many-Body Systems on a Quantum Computer},
  author = {Poulin, David and Wocjan, Pawel},
  journal = {Phys. Rev. Lett.},
  volume = {102},
  issue = {13},
  pages = {130503},
  numpages = {4},
  year = {2009},
  month = {Apr},
  publisher = {American Physical Society},
  doi = {10.1103/PhysRevLett.102.130503},
  url = {https://link.aps.org/doi/10.1103/PhysRevLett.102.130503}
}

@article{brady21b,
  author  = {Brady, Lucas T. and Baldwin, Christopher L. and Bapat, Aniruddha and Kharkov, Yaroslav and Gorshkov, Alexey V.},
  title   = {Optimal Protocols in Quantum Annealing and Quantum Approximate Optimization Algorithm Problems},
  journal = {Phys. Rev. Lett.},
  volume  = {126},
  number  = {7},
  pages   = {070505},
  year    = {2021},
  month   = feb,
  doi     = {10.1103/PhysRevLett.126.070505}
}

@article{das08,
  author  = {Das, A. and Chakrabarti, B. K.},
  title   = {Colloquium: Quantum Annealing and Analog Quantum Computation},
  journal = {Rev. Mod. Phys.},
  volume  = {80},
  number  = {3},
  pages   = {1061--1081},
  year    = {2008},
  month   = sep,
  doi     = {10.1103/RevModPhys.80.3.1061}
}

@inproceedings{gilmer2017mpnn,
  author    = {Gilmer, Justin and Schoenholz, Samuel S. and Riley, Patrick F. and Vinyals, Oriol and Dahl, George E.},
  title     = {Neural Message Passing for Quantum Chemistry},
  booktitle = {Proceedings of the 34th International Conference on Machine Learning (ICML 2017)},
  series    = {Proceedings of Machine Learning Research},
  volume    = {70},
  pages     = {1263--1272},
  year      = {2017},
  publisher = {PMLR},
  url       = {https://proceedings.mlr.press/v70/gilmer17a.html}
}

@misc{vinyals2016order,
  title         = {Order Matters: Sequence to Sequence for Sets},
  author        = {Vinyals, Oriol and Bengio, Samy and Kudlur, Manjunath},
  year          = {2015},
  eprint        = {1511.06391},
  archivePrefix = {arXiv},
  primaryClass  = {stat.ML},
  journal       = {arXiv [stat.ML]}
}

@misc{hinton2015distillation,
  title         = {Distilling the Knowledge in a Neural Network},
  author        = {Hinton, Geoffrey and Vinyals, Oriol and Dean, Jeff},
  year          = {2015},
  eprint        = {1503.02531},
  archivePrefix = {arXiv},
  primaryClass  = {cs.LG},
  journal       = {arXiv [cs.LG]},
}

@inproceedings{hu2020pretrain,
  author    = {Hu, Weihua and Liu, Bowen and Gomes, Joseph and Zitnik, Marinka and Liang, Percy and Pande, Vijay S. and Leskovec, Jure},
  title     = {Strategies for Pre-training Graph Neural Networks},
  booktitle = {8th International Conference on Learning Representations (ICLR 2020), Addis Ababa, Ethiopia, April 26--30, 2020},
  publisher = {OpenReview.net},
  year      = {2020},
  url       = {https://openreview.net/forum?id=HJlWWJSFDH}
}

@inproceedings{shi2021unimp,
  title     = {Masked Label Prediction: Unified Message Passing Model for Semi-Supervised Classification},
  author    = {Shi, Yunsheng and Huang, Zhengjie and Feng, Shikun and Zhong, Hui and Wang, Wenjing and Sun, Yu},
  booktitle = {Proceedings of the Thirtieth International Joint Conference on Artificial Intelligence (IJCAI-21)},
  publisher = {International Joint Conferences on Artificial Intelligence Organization},
  editor    = {Zhi-Hua Zhou},
  pages     = {1548--1554},
  year      = {2021},
  month     = aug,
  note      = {Main Track},
  doi       = {10.24963/ijcai.2021/214},
  url       = {https://doi.org/10.24963/ijcai.2021/214}
}

@article{malla2024,
  author  = {Malla, Rajesh K. and Sukeno, Hiroki and Yu, Hongye and Wei, Tzu-Chieh and Weichselbaum, Andreas and Konik, Robert M.},
  title   = {Feedback-based Quantum Algorithm Inspired by Counterdiabatic Driving},
  journal = {Phys. Rev. Res.},
  volume  = {6},
  number  = {4},
  pages   = {043068},
  year    = {2024},
  doi     = {10.1103/PhysRevResearch.6.043068}
}

@misc{brady2021behavioranalogquantumalgorithms,
      title={Behavior of Analog Quantum Algorithms}, 
      author={Lucas T. Brady and Lucas Kocia and Przemyslaw Bienias and Aniruddha Bapat and Yaroslav Kharkov and Alexey V. Gorshkov},
      year={2021},
      eprint={2107.01218},
      archivePrefix={arXiv},
      primaryClass={quant-ph},
  journal       = {arXiv [quant-ph]},
}

@article{egger2021warmstart,
  author  = {Egger, Daniel J. and Marecek, Jakub and Woerner, Stefan},
  title   = {Warm-starting Quantum Optimization},
  journal = {Quantum},
  volume  = {5},
  pages   = {479},
  year    = {2021},
  doi     = {10.22331/q-2021-06-17-479}
}

@misc{li2017gatedgraphsequenceneural,
  title         = {Gated Graph Sequence Neural Networks},
  author        = {Li, Yujia and Tarlow, Daniel and Brockschmidt, Marc and Zemel, Richard},
  year          = {2017},
  eprint        = {1511.05493},
  archivePrefix = {arXiv},
  primaryClass  = {cs.LG},
  journal       = {arXiv [cs.LG]},
}

@misc{khaleghian2025developmentneuralnetworkbasedoptimal,
  title         = {Development of Neural Network-Based Optimal Control Pulse Generator for Quantum Logic Gates Using the GRAPE Algorithm in NMR Quantum Computer},
  author        = {Khaleghian, Ebrahim and Fath Lipaei, Arash and Bahrampour, Abolfazl and Nikaeen, Morteza and Bahrampour, Alireza},
  year          = {2025},
  eprint        = {2412.05856},
  archivePrefix = {arXiv},
  primaryClass  = {quant-ph},
  journal       = {arXiv [quant-ph]},
}

@article{snht-7jsf,
  title = {Shadow measurements for feedback-based quantum optimization},
  author = {Bertuzzi, Let\'{\i}cia and Engster, Jo\~ao P. and da Rosa, Evandro C. R. and Duzzioni, Eduardo I.},
  journal = {Phys. Rev. A},
  volume = {112},
  issue = {2},
  pages = {022419},
  numpages = {9},
  year = {2025},
  month = {Aug},
  publisher = {American Physical Society},
  doi = {10.1103/snht-7jsf},
  url = {https://link.aps.org/doi/10.1103/snht-7jsf}
}

@article{qc91-5mj2,
  title = {Accelerating feedback-based quantum algorithms through time rescaling},
  author = {Rattighieri, L. A. M. and Pexe, G. E. L. and Bernardo, B. L. and Fanchini, F. F.},
  journal = {Phys. Rev. A},
  volume = {112},
  issue = {4},
  pages = {042607},
  numpages = {10},
  year = {2025},
  month = {Oct},
  publisher = {American Physical Society},
  doi = {10.1103/qc91-5mj2},
  url = {https://link.aps.org/doi/10.1103/qc91-5mj2}
}

@article{PhysRevResearch.7.013035,
  title = {Scalable circuit depth reduction in feedback-based quantum optimization with a quadratic approximation},
  author = {Arai, Don and Okada, Ken N. and Nakano, Yuichiro and Mitarai, Kosuke and Fujii, Keisuke},
  journal = {Phys. Rev. Res.},
  volume = {7},
  issue = {1},
  pages = {013035},
  numpages = {7},
  year = {2025},
  month = {Jan},
  publisher = {American Physical Society},
  doi = {10.1103/PhysRevResearch.7.013035},
  url = {https://link.aps.org/doi/10.1103/PhysRevResearch.7.013035}
}

@article{CLAUSEN20235171,
title = {Measurement-Based Control for Minimizing Energy Functions in Quantum Systems},
journal = {IFAC-PapersOnLine},
volume = {56},
number = {2},
pages = {5171-5178},
year = {2023},
note = {22nd IFAC World Congress},
issn = {2405-8963},
doi = {https://doi.org/10.1016/j.ifacol.2023.10.111},
url = {https://www.sciencedirect.com/science/article/pii/S2405896323004573},
author = {Henrik Glavind Clausen and Salahuddin Abdul Rahman and Özkan Karabacak and Rafal Wisniewski}
}

@article{Kosloff1,
  author  = {Kosloff, Ronnie and Hammerich, Audrey Dell and Tannor, David},
  title   = {Excitation without demolition: Radiative excitation of ground-surface vibration by impulsive stimulated Raman scattering with damage control},
  journal = {Phys. Rev. Lett.},
  volume  = {69},
  number  = {15},
  pages   = {2172--2175},
  year    = {1992},
  month   = oct,
  doi     = {10.1103/PhysRevLett.69.2172},
  url     = {https://link.aps.org/doi/10.1103/PhysRevLett.69.2172}
}

@article{Sugawara2,
  author  = {Sugawara, M. and Fujimura, Y.},
  title   = {Control of quantum dynamics by a locally optimized laser field. Application to ring puckering isomerization},
  journal = {J. Chem. Phys.},
  volume  = {100},
  number  = {8},
  pages   = {5646--5655},
  year    = {1994},
  doi     = {10.1063/1.467132}
}

@article{SUGAWARA3,
  author  = {Sugawara, M. and Fujimura, Y.},
  title   = {Control of quantum dynamics by a locally optimized laser field. Multi-photon dissociation of hydrogen fluoride},
  journal = {Chem. Phys.},
  volume  = {196},
  number  = {1},
  pages   = {113--124},
  year    = {1995},
  doi     = {10.1016/0301-0104(95)00093-4}
}

@article{OHTSUKI4,
  author  = {Ohtsuki, Y. and Yahata, Y. and Kono, H. and Fujimura, Y.},
  title   = {Application of a locally optimized control theory to pump--dump laser-driven chemical reactions},
  journal = {Chem. Phys. Lett.},
  volume  = {287},
  number  = {5},
  pages   = {627--631},
  year    = {1998},
  doi     = {10.1016/S0009-2614(98)00224-3}
}

@article{Sugawara5,
  author  = {Sugawara, M.},
  title   = {General formulation of locally designed coherent control theory for quantum system},
  journal = {J. Chem. Phys.},
  volume  = {118},
  number  = {15},
  pages   = {6784--6800},
  year    = {2003},
  month   = apr,
  doi     = {10.1063/1.1559680}
}

@article{Mirrahimi6,
  author  = {Mirrahimi, Mazyar and Turinici, Gabriel and Rouchon, Pierre},
  title   = {Reference Trajectory Tracking for Locally Designed Coherent Quantum Controls},
  journal = {J. Phys. Chem. A},
  volume  = {109},
  number  = {11},
  pages   = {2631--2637},
  year    = {2005},
  doi     = {10.1021/jp0472461},
  note    = {PMID: 16833569}
}

@article{Tannor7,
  author  = {Tannor, David J. and Kosloff, Ronnie and Bartana, Alon},
  title   = {Laser cooling of internal degrees of freedom of molecules by dynamically trapped states},
  journal = {Faraday Discuss.},
  volume  = {113},
  pages   = {365--383},
  year    = {1999},
  doi     = {10.1039/A902103E}
}

@incollection{Engel8,
  author    = {Engel, Volker and Meier, Christoph and Tannor, David J.},
  title     = {Local Control Theory: Recent Applications to Energy and Particle Transfer Processes in Molecules},
  booktitle = {Advances in Chemical Physics},
  pages     = {29--101},
  year      = {2009},
  publisher = {John Wiley \& Sons, Ltd},
  doi       = {10.1002/9780470431917.ch2}
}

@article{PhysRevLett.130.140601,
  title = {Lower Bounds on Quantum Annealing Times},
  author = {Garc\'{\i}a-Pintos, Luis Pedro and Brady, Lucas T. and Bringewatt, Jacob and Liu, Yi-Kai},
  journal = {Phys. Rev. Lett.},
  volume = {130},
  issue = {14},
  pages = {140601},
  numpages = {7},
  year = {2023},
  month = {Apr},
  publisher = {American Physical Society},
  doi = {10.1103/PhysRevLett.130.140601},
  url = {https://link.aps.org/doi/10.1103/PhysRevLett.130.140601}
}

@article{born1928beweis,
  title={Beweis des adiabatensatzes},
  author={Born, Max and Fock, Vladimir},
  journal={Z. Phys.},
  volume={51},
  number={3},
  pages={165--180},
  year={1928},
  publisher={Springer},
    url = {https://doi.org/10.1007/BF01343193}
}

@inproceedings{Grivopoulos9,
  author    = {Grivopoulos, S. and Bamieh, B.},
  title     = {Lyapunov-based control of quantum systems},
  booktitle = {42nd IEEE International Conference on Decision and Control},
  year      = {2003},
  volume    = {1},
  pages     = {434--438},
  doi       = {10.1109/CDC.2003.1272601}
}

@article{Cong1,
  author  = {Cong, Shuang and Meng, Fangfang},
  title   = {A Survey of Quantum Lyapunov Control Methods},
  journal = {Sci. World J.},
  volume  = {2013},
  pages   = {967529},
  year    = {2013},
  doi     = {10.1155/2013/967529}
}

@misc{KingmaB14,
  author        = {Kingma, Diederik P. and Ba, Jimmy},
  title         = {Adam: A Method for Stochastic Optimization},
  year          = {2015},
  eprint        = {1412.6980},
  archivePrefix = {arXiv},
  primaryClass  = {cs.LG},
}

@article{GoemansWilliamson1995,
  author  = {Goemans, Michel X. and Williamson, David P.},
  title   = {Improved approximation algorithms for maximum cut and satisfiability problems using semidefinite programming},
  journal = {J. ACM},
  volume  = {42},
  number  = {6},
  pages   = {1115--1145},
  year    = {1995},
  month   = nov,
  doi     = {10.1145/227683.227684}
}

@article{RevModPhys.94.015004,
  title = {Noisy intermediate-scale quantum algorithms},
  author = {Bharti, Kishor and Cervera-Lierta, Alba and Kyaw, Thi Ha and Haug, Tobias and Alperin-Lea, Sumner and Anand, Abhinav and Degroote, Matthias and Heimonen, Hermanni and Kottmann, Jakob S. and Menke, Tim and Mok, Wai-Keong and Sim, Sukin and Kwek, Leong-Chuan and Aspuru-Guzik, Al\'an},
  journal = {Rev. Mod. Phys.},
  volume = {94},
  issue = {1},
  pages = {015004},
  numpages = {69},
  year = {2022},
  month = {Feb},
  publisher = {American Physical Society},
  doi = {10.1103/RevModPhys.94.015004},
  url = {https://link.aps.org/doi/10.1103/RevModPhys.94.015004}
}

@article{adiabatic3,
  author  = {Albash, Tameem and Lidar, Daniel A.},
  title   = {Adiabatic quantum computation},
  journal = {Rev. Mod. Phys.},
  volume  = {90},
  number  = {1},
  pages   = {015002},
  year    = {2018},
  month   = jan,
  doi     = {10.1103/RevModPhys.90.015002}
}

@article{HamSim,
  author  = {Lloyd, Seth},
  title   = {Universal Quantum Simulators},
  journal = {Science},
  volume  = {273},
  number  = {5278},
  pages   = {1073--1078},
  year    = {1996},
  doi     = {10.1126/science.273.5278.1073}
}

@misc{wiedmann2025,
  title         = {On the convergence of the variational quantum eigensolver and quantum optimal control},
  author        = {Wiedmann, Marco and Burgarth, Daniel and Dirr, Gunther and Schulte-Herbr{\"u}ggen, Thomas and Malvetti, Emanuel and Arenz, Christian},
  year          = {2025},
  eprint        = {2509.05295},
  archivePrefix = {arXiv},
  primaryClass  = {quant-ph},
  journal       = {arXiv [quant-ph]},
}

@article{McClean_Boixo_Smelyanskiy_Babbush_Neven_2018,
  author  = {McClean, Jarrod R. and Boixo, Sergio and Smelyanskiy, Vadim N. and Babbush, Ryan and Neven, Hartmut},
  title   = {Barren Plateaus in Quantum Neural Network Training Landscapes},
  journal = {Nat. Commun.},
  volume  = {9},
  pages   = {4812},
  year    = {2018},
  month   = nov,
  doi     = {10.1038/s41467-018-07090-4}
}

@article{Cerezo_2021,
  author  = {Cerezo, M. and Sone, Akira and Volkoff, Tyler and Cincio, Lukasz and Coles, Patrick J.},
  title   = {Cost function dependent barren plateaus in shallow parametrized quantum circuits},
  journal = {Nat. Commun.},
  volume  = {12},
  pages   = {1791},
  year    = {2021},
  month   = mar,
  doi     = {10.1038/s41467-021-21728-w}
}

@article{PhysRevA.107.062421,
  title = {Optimizing quantum circuits with Riemannian gradient flow},
  author = {Wiersema, Roeland and Killoran, Nathan},
  journal = {Phys. Rev. A},
  volume = {107},
  issue = {6},
  pages = {062421},
  numpages = {11},
  year = {2023},
  month = {Jun},
  publisher = {American Physical Society},
  doi = {10.1103/PhysRevA.107.062421},
  url = {https://link.aps.org/doi/10.1103/PhysRevA.107.062421}
}

@article{x8g1-7h1k,
  title = {Nonvariational ADAPT algorithm for quantum simulations},
  author = {Tang, Ho Lun and Chen, Yanzhu and Biswas, Prakriti and Magann, Alicia B. and Arenz, Christian and Economou, Sophia E.},
  journal = {Phys. Rev. Res.},
  volume = {7},
  issue = {2},
  pages = {023275},
  numpages = {15},
  year = {2025},
  month = {Jun},
  publisher = {American Physical Society},
  doi = {10.1103/x8g1-7h1k},
  url = {https://link.aps.org/doi/10.1103/x8g1-7h1k}
}

@article{PhysRevResearch.5.033227,
  title = {Randomized adaptive quantum state preparation},
  author = {Magann, Alicia B. and Economou, Sophia E. and Arenz, Christian},
  journal = {Phys. Rev. Res.},
  volume = {5},
  issue = {3},
  pages = {033227},
  numpages = {8},
  year = {2023},
  month = {Sep},
  publisher = {American Physical Society},
  doi = {10.1103/PhysRevResearch.5.033227},
  url = {https://link.aps.org/doi/10.1103/PhysRevResearch.5.033227}
}

@misc{malvetti2025,
  title         = {Randomized Gradient Descents on Riemannian Manifolds: Almost Sure Convergence to Global Minima in and beyond Quantum Optimization},
  author        = {Malvetti, Emanuel and Arenz, Christian and Dirr, Gunther and Schulte-Herbr{\"u}ggen, Thomas},
  year          = {2025},
  eprint        = {2405.12039},
  archivePrefix = {arXiv},
  primaryClass  = {math.OC},
  journal       = {arXiv [math.OC]},
}

@misc{mcmahon2025,
  title         = {Equating quantum imaginary time evolution, Riemannian gradient flows, and stochastic implementations},
  author        = {McMahon, Nathan A. and Pervez, Mahum and Arenz, Christian},
  year          = {2025},
  eprint        = {2504.06123},
  archivePrefix = {arXiv},
  primaryClass  = {quant-ph},
  journal       = {arXiv [quant-ph]}}

@article{trotter1959,
  author  = {Trotter, H. F.},
  title   = {On the Product of Semi-Groups of Operators},
  journal = {Proc. Am. Math. Soc.},
  volume  = {10},
  number  = {4},
  pages   = {545--551},
  year    = {1959},
  doi     = {10.1090/S0002-9939-1959-0108732-6}
}

@article{suzuki1976,
  author  = {Suzuki, M.},
  title   = {Generalized Trotter's Formula and Systematic Approximants of Exponential Operators and Inner Derivations with Applications to Many Body Problems},
  journal = {Commun. Math. Phys.},
  volume  = {51},
  pages   = {183--190},
  year    = {1976},
  doi     = {10.1007/BF01609348}
}

@article{domain1,
  author  = {Pan, Sinno Jialin and Yang, Qiang},
  title   = {A Survey on Transfer Learning},
  journal = {IEEE Trans. Knowl. Data Eng.},
  year    = {2010},
  volume  = {22},
  number  = {10},
  pages   = {1345--1359},
  doi     = {10.1109/TKDE.2009.191}
}

@article{domain2,
  author  = {Ganin, Yaroslav and Ustinova, Evgeniya and Ajakan, Hana and Germain, Pascal and Larochelle, Hugo and Laviolette, Fran{\c{c}}ois and Marchand, Mario and Lempitsky, Victor},
  title   = {Domain-Adversarial Training of Neural Networks},
  journal = {J. Mach. Learn. Res.},
  year    = {2016},
  volume  = {17},
  number  = {59},
  pages   = {1--35},
  url     = {http://jmlr.org/papers/v17/15-239.html}
}

@inproceedings{domain3,
  author    = {Tzeng, Eric and Hoffman, Judy and Saenko, Kate and Darrell, Trevor},
  title     = {Adversarial Discriminative Domain Adaptation},
  booktitle = {2017 IEEE Conference on Computer Vision and Pattern Recognition (CVPR)},
  year      = {2017},
  pages     = {2962--2971},
  doi       = {10.1109/CVPR.2017.316}
}

@inproceedings{meta1,
  author    = {Finn, Chelsea and Abbeel, Pieter and Levine, Sergey},
  title     = {Model-Agnostic Meta-Learning for Fast Adaptation of Deep Networks},
  booktitle = {Proceedings of the 34th International Conference on Machine Learning (ICML)},
  series    = {Proceedings of Machine Learning Research},
  volume    = {70},
  pages     = {1126--1135},
  year      = {2017},
  publisher = {PMLR},
  url       = {https://proceedings.mlr.press/v70/finn17a.html}
}

@article{meta2,
  author  = {Hospedales, Timothy and Antoniou, Antreas and Micaelli, Paul and Storkey, Amos},
  title   = {Meta-Learning in Neural Networks: A Survey},
  journal = {IEEE Trans. Pattern Anal. Mach. Intell.},
  year    = {2022},
  volume  = {44},
  number  = {9},
  pages   = {5149--5169},
  doi     = {10.1109/TPAMI.2021.3079209}
}

@article{garcia2025tighter,
  author  = {Garc{\'\i}a-Pintos, Luis Pedro and Sahasrabudhe, Mrunmay and Arenz, Christian},
  title   = {Tighter lower bounds on quantum annealing times},
  journal = {SciPost Phys.},
  volume  = {18},
  number  = {5},
  pages   = {159},
  year    = {2025},
  doi     = { 10.21468/SciPostPhys.18.5.159}
}

@article{Abdul_Rahman_2026,
  author  = {Abdul Rahman, Salahuddin and Karabacak, {\"O}zkan and Wisniewski, Rafal},
  title   = {Feedback-based quantum strategies for constrained combinatorial optimization problems},
  journal = {Future Gener. Comput. Syst.},
  volume  = {174},
  pages   = {107979},
  year    = {2026},
  doi     = {10.1016/j.future.2025.107979}
}

@inproceedings{Rahman2024,
  author    = {Abdul Rahman, Salahuddin and Karabacak, Ozkan and Wisniewski, Rafal},
  title     = {Weighted Feedback-Based Quantum Algorithm for Excited States Calculation},
  booktitle = {2024 IEEE International Conference on Quantum Computing and Engineering (QCE)},
  year      = {2024},
  pages     = {169--175},
  month     = sep,
  doi       = {10.1109/QCE60285.2024.00029}
}

@article{roland2002quantum,
  author  = {Roland, J{\'e}r{\'e}mie and Cerf, Nicolas J.},
  title   = {Quantum search by local adiabatic evolution},
  journal = {Phys. Rev. A},
  volume  = {65},
  pages   = {042308},
  year    = {2002},
  doi     = {10.1103/PhysRevA.65.042308}
}

@article{kolodrubetz2017geometry,
  author  = {Kolodrubetz, Michael and Sels, Dries and Mehta, Pankaj and Polkovnikov, Anatoli},
  title   = {Geometry and non-adiabatic response in quantum and classical systems},
  journal = {Phys. Rep.},
  volume  = {697},
  pages   = {1--87},
  year    = {2017},
  doi     = {10.1016/j.physrep.2017.07.001}
}

@article{sels2017minimizing,
  author  = {Sels, Dries and Polkovnikov, Anatoli},
  title   = {Minimizing irreversible losses in quantum systems by local counterdiabatic driving},
  journal = {Proc. Natl. Acad. Sci. U.S.A.},
  volume  = {114},
  number  = {20},
  pages   = {E3909--E3916},
  year    = {2017},
  doi     = {10.1073/pnas.1619826114}
}

@article{gueryodelin2019shortcuts,
  author  = {Gu{\'e}ry-Odelin, David and Ruschhaupt, Andreas and Kiely, Anthony and Torrontegui, Erik and Mart{\'i}nez-Garaot, Sofia and Muga, Juan Gonzalo},
  title   = {Shortcuts to adiabaticity: Concepts, methods, and applications},
  journal = {Rev. Mod. Phys.},
  volume  = {91},
  pages   = {045001},
  year    = {2019},
  doi     = {10.1103/RevModPhys.91.045001}
}

@article{claeys2019floquet,
  author = {Claeys, Pieter W. and Pandey, Mohit and Sels, Dries and Polkovnikov, Anatoli},
  title = {Floquet-Engineering Counterdiabatic Protocols in Quantum Many-Body Systems},
  journal = {Phys. Rev. Lett.},
  volume = {123},
  pages = {090602},
  year = {2019},
  doi = {10.1103/PhysRevLett.123.090602}
}
\end{document}